\documentclass[letterpaper,11pt]{article}
\pdfoutput=1

\usepackage{jheppub}
\usepackage{multirow}
\usepackage{shuffle}

\usepackage{subfig}
\usepackage{xspace}
\usepackage[countmax]{subfloat}
\usepackage{enumitem}
\usepackage{amssymb}
\usepackage{amsmath}
\usepackage{cancel}
\usepackage{color}
\usepackage{braket}
\usepackage{graphicx}
\usepackage{multirow}
\usepackage{verbatim}
\usepackage{amsthm}
\usepackage{slashed}
\usepackage{wasysym}
\usepackage{simplewick}
\usepackage{mathtools}
\usepackage{soul}
\usepackage{xspace}

\newcommand{\mcdot}{\!\cdot\!}

\newcommand{\gcusp}{\Gamma^{\mathrm{cusp}}}
\def\Ord{{\cal O}}

\def\cG{\mathcal{G}}

\def\cI{\mathcal{I}}

\def\cL{\mathcal{L}}

\def\cN{\mathcal{N}}
\def\cO{\mathcal{O}}

\def\cP{\mathcal{P}}

\newcommand{\brk}{\nonumber\\&}
\newcommand{\nbrk}{\nonumber\\}

\def\ugr{\gamma^{r}}

\def\Lp{L_\perp}

\def\nn{{\nonumber}}
\def\mcdot{\!\cdot\!}

\newcommand{\EEC}{{\text{EEC}}}

\newcommand{\gsoft}{\gamma^{s}}

\newcommand{\Eq}[1]{Equation~\eqref{#1}}

\DeclareRobustCommand{\Sec}[1]{Sec.~\ref{#1}}

\DeclareRobustCommand{\App}[1]{App.~\ref{#1}}

\DeclareRobustCommand{\Fig}[1]{Fig.~\ref{#1}}

\DeclareRobustCommand{\Eq}[1]{Eq.~(\ref{#1})}
\DeclareRobustCommand{\Eqs}[2]{Eqs.~(\ref{#1}) and (\ref{#2})}

\def\be{\begin{equation}}
\def\ee{\end{equation}}

\newcommand{\SCETi}{\mbox{${\rm SCET}_{\rm I}$}\xspace}
\newcommand{\SCETii}{\mbox{${\rm SCET}_{\rm II}$}\xspace}

\allowdisplaybreaks[3]

\newcommand{\df}{\mathrm{d}}

\newcommand{\e}{\epsilon}

\newcommand{\zero}{{(0)}}
\newcommand{\one}{{(1)}}

  \newcount\hour \newcount\minute
  \hour=\time \divide \hour by 60 \minute=\time
  \newcommand{\todaytime}{\today \ -- \number\hour :\ifnum \minute<10 0\fi\number\minute}

\title{Simplicity from Recoil: \\The Three-Loop Soft Function and\\ Factorization for the Energy-Energy Correlation}

\author[1,2]{Ian Moult}
\author[3]{and Hua Xing Zhu}

\affiliation[1]{Berkeley Center for Theoretical Physics, University of California, Berkeley, CA 94720, USA}
\affiliation[2]{Theoretical Physics Group, Lawrence Berkeley National Laboratory, Berkeley, CA 94720, USA}
\affiliation[3]{Zhejiang Institute of Modern Physics, Department of Physics, Zhejiang University, Hangzhou, Zhejiang 310027, China}

\emailAdd{ianmoult@lbl.gov}
\emailAdd{zhuhx@zju.edu.cn}

\abstract{We derive an operator based factorization theorem for the energy-energy correlation (EEC) observable in the back-to-back region, allowing the cross section to be written as a convolution of hard, jet and soft functions. We prove the equivalence of the soft functions for the EEC and color singlet transverse-momentum resummation to all-loop order, and give their analytic result to three-loops. Large logarithms appearing in the perturbative expansion of the EEC can be resummed to all orders using renormalization group evolution. We give analytic results for all required anomalous dimensions to three-loop order, providing the first example of a transverse-momentum (recoil) sensitive $e^+e^-$ event shape whose anomalous dimensions are known at this order. The EEC can now be computed to next-to-next-to-next-to-leading logarithm matched to next-to-next-to-leading order, making it a prime candidate for precision QCD studies and extractions of the strong coupling constant. We anticipate that our factorization theorem will also be crucial for understanding non-perturbative power corrections for the EEC, and their relationship to those appearing in other observables. 
}

\begin{document} 

\maketitle

\section{Introduction} \label{sec:intro}

Event shapes in $e^+e^-$ provide a theoretically and experimentally clean environment for precision studies of QCD and extractions of the strong coupling constant, $\alpha_s$.  The perturbative description of event shapes requires both the calculation of fixed order corrections, which are currently known to next-to-next-to-leading order~(NNLO) \cite{GehrmannDeRidder:2007hr,Gehrmann-DeRidder:2007nzq,Weinzierl:2008iv,Weinzierl:2009ms}, as well as the all orders resummation of singular logarithmic terms \cite{Catani:1992ua}. There has been significant progress in both of these areas in recent years: advances in fixed order subtractions \cite{DelDuca:2016csb,DelDuca:2016ily,Somogyi:2006db,Somogyi:2006da}, have enabled a number of new NNLO calculations \cite{DelDuca:2016csb,DelDuca:2016ily}, and resummation to next-to-next-to-leading logarithmic (NNLL) order has been performed for a wide variety of observables \cite{deFlorian:2004mp,Becher:2012qc,Banfi:2016zlc,Frye:2016okc,Frye:2016aiz,Mo:2017gzp,Tulipant:2017ybb}, and implemented in a fully generic manner \cite{Banfi:2004yd,Banfi:2014sua}.

Resummation to next-to-next-to-next-to-leading logarithmic (N$^3$LL) accuracy has been achieved using the soft collinear effective theory (SCET) \cite{Bauer:2000ew, Bauer:2000yr, Bauer:2001ct, Bauer:2001yt}, which allows resummation to be performed using renormalization group evolution, in virtuality or rapidity \cite{Chiu:2011qc,Becher:2011dz,Chiu:2012ir}, of field theoretic operators. Resummed predictions at N$^3$LL accuracy have been made for thrust \cite{Becher:2008cf,Abbate:2010xh}, the $C$-parameter \cite{Hoang:2014wka} and heavy jet mass \cite{Chien:2010kc}, enabling precision extractions of $\alpha_s$ \cite{Becher:2008cf,Chien:2010kc,Abbate:2010xh,Hoang:2015hka}. However, all of these observables are ``recoil free", or transverse-momentum ($q_T$) insensitive (described by $\SCETi$), meaning that at leading power, soft partons are not able to recoil the $q_T$ of the jet. Unfortunately, there are no examples of $q_T$ sensitive observables in $e^+e^-$ which are known at N$^3$LL accuracy, which could complement $\alpha_s$ fits using recoil free observables. Recoil typically significantly complicates perturbative calculations. For example, calculations of jet broadening \cite{Rakow:1981qn,Catani:1992jc,Dokshitzer:1998kz} are complicated by issues of recoil, jet regions, and by the fact that it is a scalar sum which become complicated at multiple emissions. Indeed, elliptic functions appear already at NNLL for broadening \cite{Becher:2012qc}, making extensions to higher orders seem difficult. 

Recently, the three-loop soft function governing the color singlet $q_T$ spectrum at small $q_T$ was computed \cite{Li:2016ctv}. This calculation used bootstrap techniques from $\cN=4$ super Yang-Mills theory \cite{Dixon:2011pw,Dixon:2014iba,Dixon:2015iva,Caron-Huot:2016owq,Dixon:2016nkn}, a supersymmetric decomposition in transcendental weight, a newly introduced rapidity regulator \cite{Li:2016axz}, multi-dimensional factorization \cite{Larkoski:2014tva,Procura:2014cba}, and recently computed master integrals \cite{Duhr:2013msa,Li:2013lsa,Anastasiou:2013srw,Li:2014bfa,Zhu:2014fma,Anastasiou:2015yha}. The final result has a remarkably simple structure and exhibits interesting relations to other anomalous dimensions \cite{Li:2016ctv,Vladimirov:2016dll,Vladimirov:2017ksc}. The computation to this order was ultimately enabled by the simple structure of the observable: it is a vector sum, which preserves the maximal number of rotational symmetries, and does not involve any jet regions, or projections onto axes whose precise definition can modify the perturbative structure \cite{Larkoski:2014uqa,Larkoski:2015uaa}.  It is therefore interesting to ask whether this anomalous dimension controls the resummation of any $e^+e^-$ event shape observables. This is interesting both phenomenologically, as it could provide information for $\alpha_s$ extractions complimentary to that from recoil free observables, as well as for understanding all orders relations between different observables.

In this paper, we derive an all orders factorization theorem for the energy-energy correlation (EEC) in the back-to-back region. In this factorization theorem, the soft radiation does not contribute directly to the observable at leading power, but instead contributes only via recoil. We are able to show that the soft function appearing in the factorization is identical to that for the color singlet $q_T$ distribution, up to the direction of the Wilson lines.\footnote{The similarity between the resummation for EEC and $q_T$ in the back-to-back region has long been known, and has been used to perform the resummation to NNLL using the Collins-Soper-Sterman form as an ansatz, see e.g.  \cite{deFlorian:2004mp,Tulipant:2017ybb}. However, we were not able to find a factorization theorem for the EEC in the literature, or an all orders proof of this relation between the anomalous dimensions. The steps for a proof of factorization along with a leading log resummation formula were given in \cite{Collins:1981uk}.} Using a recently introduced rapidity regulator \cite{Li:2016axz}, which allows both the regulator and the measurement function to be described by spacetime shifts of the Wilson lines, we prove that the soft function is invariant under the crossing of the Wilson lines, allowing us to use the recently derived results for $q_T$ soft function to derive the anomalous dimension and soft function for the EEC. This provides the first example of a $q_T$ sensitive observable in $e^+e^-$ whose anomalous dimensions are known to three-loops. It also illustrates the utility of operator definitions in factorization theorems, which allow for the identification of universal structures in apparently different situations. As a further consequence of our analysis, the anomalous dimension and soft function could also be used for identified hadron production in the back-to-back limit, allowing it to be extended to N$^3$LL perturbative accuracy.

\begin{figure}
  \centering
  \includegraphics[width=0.5\textwidth]{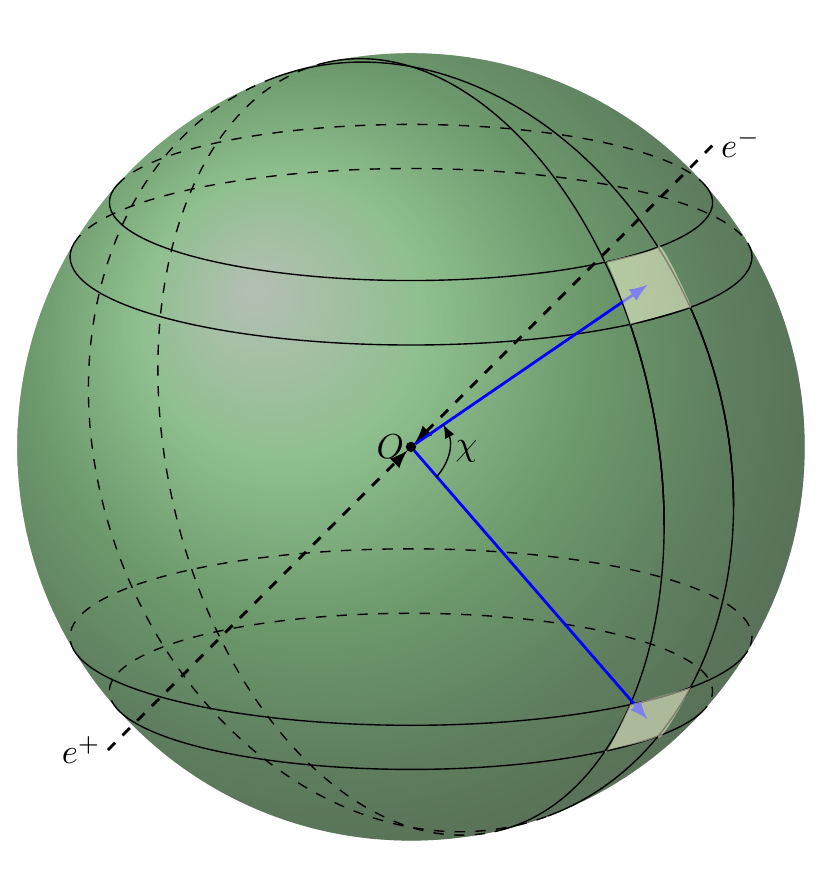}
  \caption{An illustration of the EEC observable in $e^+e^-$ annihilation, which is defined as the energy-energy correlation of two calorimeter cells with opening angle $\chi$.}
  \label{fig:EEC_def}
\end{figure}

The EEC, one of the earliest examples of an infrared and collinear (IRC) safe observable, is defined as \cite{Basham:1978bw}
\begin{align}\label{eq:EEC_intro}
\text{EEC}=\sum\limits_{a,b} \int d\sigma_{V\to a+b+X} \frac{2 E_a E_b}{Q^2 \sigma_{\text{tot}}}   \delta(\cos(\theta_{ab}) - \cos(\chi))\,,
\end{align}
where the sum is over all \emph{different} pairs of hadrons $h_a$ and
$h_b$ in the event, c.f. Fig.~\ref{fig:EEC_def}. 
It has been studied extensively in the QCD literature \cite{Ali:1982ub,Chao:1982wb,Richards:1982te,Ellis:1983fg,Schneider:1983iu,Richards:1983sr,Falck:1988gb,Fiore:1992sa,Glover:1994vz,Clay:1995sd,Kramer:1996qr,Dokshitzer:1999sh,deFlorian:2004mp}, and has been computed analytically to NLO in planar $\cN=4$ super Yang-Mills theory, exploiting a relation to correlation functions \cite{Belitsky:2013xxa,Belitsky:2013bja,Belitsky:2013ofa}, as well as at strong-coupling \cite{Hofman:2008ar} using the AdS/CFT correspondence \cite{Maldacena:1997re}. There has also been progress towards the NLO calculation in QCD \cite{Gituliar:2017aa}.
Recently it was computed at NNLL+NNLO \cite{Tulipant:2017ybb} using the NNLO calculation of \cite{DelDuca:2016csb,DelDuca:2016ily} and used to fit $\alpha_s$ from data. Our results will allow this to be extended to N$^3$LL'+NNLO, matching the state of the art precision for recoil free ($\SCETi$) observables.\footnote{Here the prime on the logarithmic accuracy indicates the inclusion of the three-loop boundary conditions for the soft and collinear functions, as has been included for thrust and $C$-parameter \cite{Becher:2008cf,Abbate:2010xh,Hoang:2014wka}.  See also \cite{Almeida:2014uva} for a detailed discussion of order counting.}

An outline of this paper is as follows. In \Sec{sec:kinematics} we discuss the kinematics of the EEC in the back-to-back limit, and illustrate the relationship between the EEC and $q_T$. In \Sec{sec:fact} we present our factorization theorem for the EEC observable, discussing in detail how soft radiation contributes to the observable.  In \Sec{sec:univ} we prove that both the anomalous dimensions, as well as the full soft function, are identical to those governing the color singlet $q_T$ spectrum, and we use this relation to give explicit results for the soft function to three-loop order. In \Sec{sec:resum} we use our factorization theorem to give the all orders form for the resummed cross section in the back-to-back limit. We conclude and discuss future directions in \Sec{sec:conclusions}. Additional calculations, and a comparison to NNLO of the logarithmic structure as predicted by our factorization theorem with results in the literature, are given in the appendices.

\section{Kinematics} \label{sec:kinematics}

In this section we discuss in detail the kinematics of the EEC observable in the back-to-back region, $\chi\to \pi$ in \Eq{eq:EEC_intro}. This will be important in understanding the derivation of the factorization theorem, and the relation to the $q_T$ observable. It will be convenient to work with the dimensionless variable
\begin{align}
z=\frac{1-\cos \theta_{ij}}{2}\,,
\end{align}
in terms of which we have
\begin{align}\label{eq:eec_defz}
\frac{d\sigma}{dz} = \frac{1}{2} \sum\limits_{ij} \int dx_i dx_j  x_i x_j \frac{d^3 \sigma}{dx_i dx_j dz}\,,
\end{align}
where $\frac{d^3 \sigma}{dx_i dx_j dz}$ is the triple differential
cross section measuring the energy fraction with respect to half of
the center of mass energy, $x_{i,j} = 2 E_{i,j}/Q$, and
relative angle $z$. Note that this triple differential distribution
is not IRC safe. IRC safety is recovered after summing over different
particles. Here the summation is over different pair of hadron in the final state with momentum $p_i$ and
$p_j$.
The back-to-back limit is then characterized by $\chi = \theta_{ij} \to \pi$, or $z\to 1$.

\begin{figure}
\begin{center}
\includegraphics[width=0.75\columnwidth]{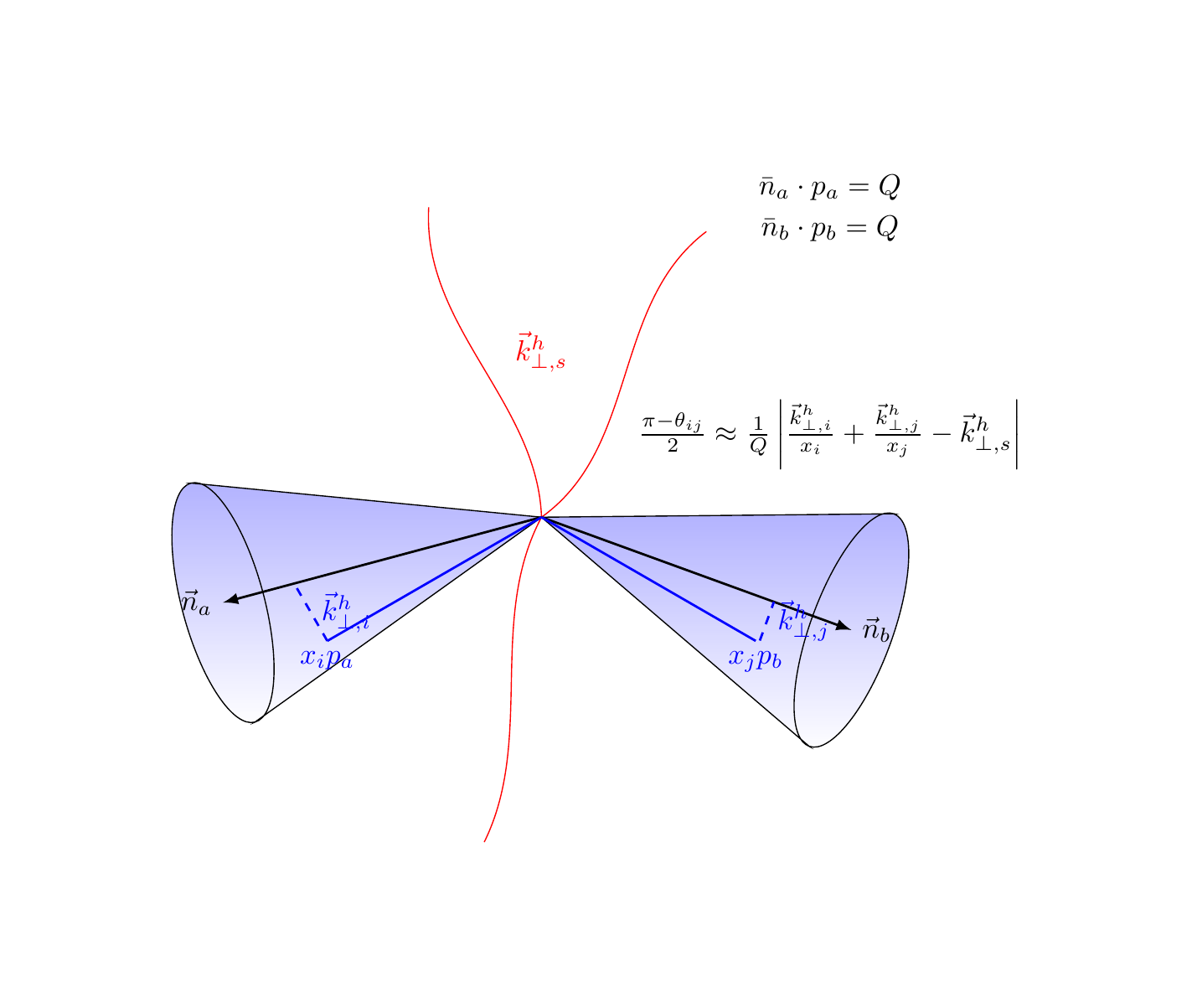}
\end{center}
\caption{A schematic of the kinematics relevant for the description of the EEC observable in the back-to-back limit, $z\to 1$. The total transverse momentum of final-state hadrons perpendicular to the thrust axis~(the black dashed line) is denoted as $\vec{k}_{\perp,s}^h$. The three-momentum of a pair of collinear hadrons which enter weighted sum in \Eq{eq:eec_defz} are denoted as $\vec{k}_{i,j}^h$. The corresponding transverse components perpendicular to the jet axis $\vec{n}_{a,b}$ are $\vec{k}_{\perp,i}^h$ and $\vec{k}_{\perp,j}^h$. Soft radiation acts only to recoil the two collinear sectors, denoted by $\vec n_{a,b}$, with respect to each other, but does not contribute directly to the observable.}
\label{fig:schematic_factorization}
\end{figure}

In the back-to-back limit, the event consists of two nearly back-to-back jets, along with additional low energy (soft) radiation. Additional hard jets are power suppressed (for a detailed discussion, see \cite{Chiu:2012ir}). This situation is illustrated schematically in \Fig{fig:schematic_factorization}. We denote the momentum of the two jets by $p_a^\mu=(p_a^0,\vec{p}_a)$ and $p_b^\mu = (p_b^0, \vec{p}_b)$. We then define two \emph{light-cone} directions $n_a^\mu = (1, \vec{n}_a)$ and $n_b^\mu = (1, \vec{n}_b)$, with $\vec{n}_{a(b)}=\vec{p}_{a(b)}/|\vec{p}_{a(b)}|$. We also define the conjugate directions $\bar{n}_{a(b)} = (1, - \vec{n}_{a(b)})$.
The relevant modes in the effective theory are easily determined by considering on-shell modes that contribute to the EEC observable at leading power. They are found to be soft, collinear, and anti-collinear, with the scalings in light-cone coordinates
\begin{align}\label{eq:pc_modes}
p_s\sim Q(\lambda, \lambda, \lambda)\,, \qquad p_c \sim Q(\lambda^2, 1, \lambda) \,, \qquad p_{\bar c} \sim Q(1,\lambda^2,  \lambda)\,,
\end{align}
where
\begin{align}
\lambda\sim \sqrt{1-z}\,.
\end{align}
In particular, we see that the EEC is an SCET$_{\text{II}}$ \cite{Bauer:2002aj} observable. This is intuitively obvious, since the EEC directly measures angles between hadrons and is therefore sensitive to recoil at leading power. With the above definition, we have $\bar{n}_{a(b)} \mcdot p_{a(b)} = Q + \Ord(\lambda Q)$.

Using the definition of the observable in \Eq{eq:eec_defz}, and the power counting of the modes in \Eq{eq:pc_modes}, we can now expand the EEC observable to leading power in the $z\to 1$ limit.  We begin by noting that the contribution of soft modes to the observable is power suppressed. Soft radiation therefore will contribute only by recoiling the jet sectors. This is quite interesting, and in particular, it implies that it is sufficient to know the total vector transverse momentum of the soft sector.\footnote{We find it interesting that recoil, which often leads to complications, in fact leads to the remarkable simplicity of the soft function for the EEC.} 

Contributions to the observable arise only from correlations between collinear partons in different collinear sectors.
It is therefore a simple geometric exercise to relate their perpendicular momentum to $z$, as relevant for the EEC. Considering the configuration shown in \Fig{fig:schematic_factorization}, 
to leading power we find
\begin{align}
\label{eq:z_in_back_to_back}
1-z=\frac{1}{Q^2}  \left|  \frac{\vec k_{\perp,i}^h}{x_i}+\frac{\vec k_{\perp,j}^h}{x_j}-\vec k_{\perp,s}^h \right|^2 +\cO(1-z)\,,
\end{align}
where $k_{\perp,s}^h$ is the total transverse momentum of soft final-state hadrons
relative to thrust axis, and $k_{\perp,i(j)}^h$ is the tranverse momentum of a collinear hadron relative to its respective jet axis, defined as the direction with largest energy flow. We emphasize that \Eq{eq:z_in_back_to_back} only holds in the back-to-back limit. With these definitions, we have the conservation of transverse momentum within each jet
\begin{align}
  \label{eq:jet_pt_conservation}
  \sum_{i \in \text{jet a}} \vec{k}_{\perp,i}^h = \Ord(\lambda^2 Q) \,,
\qquad
  \sum_{j \in \text{jet b}} \vec{k}_{\perp,j}^h = \Ord(\lambda^2 Q) \,,
\end{align}
and the conservation of total transverse momentum perpendicular to the thrust axis
\begin{align}
  \label{eq:total_pt_conservation}
  \vec{k}_{\perp,s}^h + \vec{p}_{\perp,a} + \vec{p}_{\perp,b} = \Ord(\lambda^2 Q) \,,
\end{align}
where $\vec{p}_{\perp,a(b)}$ is the transverse momentum of jet a(b) perpendicular to the thrust axis. 
\Eq{eq:z_in_back_to_back} clearly shows the relationship between the EEC observable and $q_T$. This expression also hints at the simplicity of the EEC. Most importantly, $1-z$ is related to the vector sum of the transverse-momentum in the different sectors, and in particular, the only property of the soft radiation that is measured is the total transverse-momentum. This is much simpler than other recoil sensitive $e^+e^-$ observables such as broadening, where it is ultimately the scalar sum of the transverse momentum that is measured, making the measurement function extremely complicated for configurations with multiple emissions.

\section{Factorization Theorem for the EEC in the Back-to-Back Region} \label{sec:fact}

Having understood the kinematics of the EEC observable in the back-to-back region, we can now derive a factorization theorem describing the all orders singular behavior in the  $z\to1$ ($\chi\to \pi$) limit. This factorization theorem will allow us to express the leading power cross section for the EEC as a product (convolution) of a hard matching coefficient, $H$, a soft function, $S$, which describes wide angle soft radiation, and jet functions, $J$, which describe collinear radiation in the jets. Each of these functions will describe the dynamics at a single scale, and large logarithms in the cross section can be resummed through their renormalization group evolution, which is given in \Sec{sec:RG}.

We will present this factorization in the language of SCET \cite{Bauer:2000ew, Bauer:2000yr, Bauer:2001ct, Bauer:2001yt}, giving gauge invariant operator definitions for the jet and soft functions. While the resummation of the EEC to NNLL order has been performed in the Collins-Soper-Sterman (CSS) \cite{Collins:1981uk,Collins:1981va,Kodaira:1981nh,Kodaira:1982az,Collins:1984kg} formalism (see e.g. \cite{deFlorian:2004mp,Tulipant:2017ybb}), we are not aware of a factorization theorem in terms of hard, jet and soft functions. Such a factorization will ultimately allow us to prove the equivalence of the soft function for the EEC and for $q_T$ in color singlet production, and to extend the perturbative accuracy of the EEC observable to N$^3$LL.

\subsection{Factorization Theorem} \label{sec:fact_th}

We will derive the factorization theorem for the case of $e^+e^-\to$ dijets through an off-shell photon or $Z$. The extension to other underlying hard processes, such as $e^+e^-\to gg$ through an off-shell Higgs, is trivial. Our proof will use as a starting point the factorization theorem for identified hadron production in the back-to-back limit \cite{Collins:1981uk,Collins:1981va}, and we will formulate the factorization for the EEC observable by marginalizing over this factorization theorem. Since a factorization theorem, and in particular the cancellation of Glauber modes, has been proven for back-to-back identified hadron production, we will also be able to use this argument to conclude that Glauber modes do not contribute to the EEC and therefore that they do not violate the factorization at all orders.

Since the derivation of the factorization starts from identified hadron production in the back-to-back limits, we will use fragmentation functions, and their transverse momentum dependent (TMD) counterparts, extensively. We therefore begin by reviewing their definitions.
We define the standard fragmentation functions (FFs) as \cite{Georgi:1977mg,Ellis:1978ty,Collins:1981uw,Collins:1989gx}
\begin{align}
f_{q\to h} (z_h) &=\frac{1}{4 z_h N_c} \sum\limits_X \int \frac{d\xi^+}{4\pi} e^{-i p_h^- \xi^+/z_h} \\
& \hspace{3cm}\langle 0 | T \left[ \tilde W^\dagger_{n} q_j  \right]_a \left( \frac{\xi^+}{2} \right) |X,h \rangle \gamma_{ij}^- \langle X,h| \bar T \left[  \bar q_i \tilde W_{n} \right]_a \left( -\frac{\xi^+}{2} \right ) |0 \rangle \,,\nn
\end{align}
and their TMD extensions (TMDFFs) as \cite{Collins:2011zzd,Echevarria:2016scs}
\begin{align}\label{eq:TMDFF_def}
F_{q\to h} (\vec b_\perp,z_h) &=\frac{1}{4 z_h N_c} \sum\limits_X \int \frac{d\xi^+}{4\pi} e^{-i p_h^- \xi^+/z_h} \\
&\hspace{3cm} \langle 0 | T \left[ \tilde W^\dagger_{n} q_j  \right]_a \left( \frac{\xi}{2} \right) |X,h \rangle \gamma_{ij}^- \langle X,h| \bar T \left[  \bar q_i \tilde W_{n} \right]_a \left( -\frac{\xi}{2} \right ) |0 \rangle\,.\nn
\end{align}
Here $W_n$ is a Wilson line, defined in momentum space as
\begin{align}
W_n =\left[ \, \sum\limits_{\text{perms}} \exp \left(  -\frac{g}{\bar{n}\cdot \cP} \bar n \cdot A_n(x)  \right) \right]\,,
\end{align}
and $q$ are lightcone projected fermionic fields. Here $\xi=(\xi^+,0^-,\vec b_\perp)$, with $\vec b_\perp$ the conjugate variable to $\vec k_\perp^h$, the transverse momentum of $h$ perpendicular to jet axis $\vec{n}$, which is aligned with the total jet three momentum. The lightcone vector $n^\mu$ in the operator definition of $f_{q\to h}$ and $F_{q \to h}$ is then defined by $n^\mu = (1, \vec{n})$. We also define the the conjugate lightcone vector $\bar{n}^\mu = (1, - \vec{n})$.   We are therefore working in the center-of-mass frame, in contrast to the more conventional hadron frame~\cite{Collins:2011zzd}. 
Note that the only difference between the definition for the TMDFF and the standard FF is in the positions of the fields. The renormalized fragmentation functions satisfy the following sum rule
\begin{align}\label{eq:sum_rule}
\sum\limits_h \int \limits_0^1 dx~x~ f_{q\rightarrow h}(x,\mu_F)=1\,,
\end{align}
which will play an important role in our derivation.

To derive a factorization theorem for the EEC in the back-to-back region, we begin by factorizing the multi-differential cross section which appears in its definition
\begin{align}
\frac{d\sigma}{dz} = \frac{1}{2} \sum\limits_{ij} \int dx_i dx_j  x_i x_j \frac{d^3 \sigma}{dx_i dx_j dz}\,.
\end{align}
Furthermore, in the back-to-back limit, we can exchange the variable $z$ for an auxiliary transverse momentum, $\vec k_\perp = \vec{k}_{\perp,i}^h/x_i + \vec{k}_{\perp,j}^h/x_j - \vec{k}_{\perp,s}^h$, by writing
\begin{align}
\frac{d^3 \sigma}{dx_i dx_j dz} = \int d^2 \vec k_\perp  \frac{d^3 \sigma}{dx_i dx_j d^2 \vec k_\perp} \delta \left(  1-z -\frac{\vec k_\perp^2}{Q^2}  \right)\,.
\end{align}
In the back-to-back limit we can write a factorized expression for this cross section using the result for identified hadrons in the back-to-back region as was studied in the seminal papers \cite{Collins:1981va,Collins:1981uk}. In \cite{Collins:1981va,Collins:1981uk} a factorization theorem was proven, and in particular it was shown that Glauber modes do not contribute, using techniques developed in \cite{Sterman:1978bi,Collins:1981ta,Collins:1988ig} (see also \cite{Collins:1989gx,Collins:2011zzd} for a review). Since we will formulate our factorization for the EEC from this starting point, this implies also that Glaubers cancel from the EEC observable. Using these results, we have
\begin{align}
\frac{d^3 \sigma}{dx_i dx_j d^2 \vec k_\perp} =& H(Q,\mu)    \int d^2 \vec k_{\perp,i}^h \int d^2 \vec k_{\perp,j}^h \int d^2 \vec k_{\perp,s} ~ \delta^{(2)} \left(  \vec k_\perp -\left(    \frac{\vec k_{\perp,i}^h}{x_i}+\frac{\vec k_{\perp,j}^h}{x_j}-\vec k_{\perp,s} \right)  \right) \nn \\
&\cdot F_{q\to i}(\vec k_{\perp,i}^h, x_i,\mu,\nu)  F_{q\to j}(\vec k_{\perp,j}^h, x_j,\mu,\nu)  S_\EEC(\vec k_{\perp,s},\mu,\nu)\,,
\end{align}
%
where the triple differential distribution is written as convolution over transverse momentum of collinear hadrons perpendicular to jet axis, and transverse momentum of soft hadrons perpendicular to thrust axis.
For the soft sector we do not distinguish partonic and hadronic momentum, as the impact of soft modes to the factorization formula is only through recoil.
Here $H(Q,\mu)$ is the hard matching coefficient for $e^+e^-\to q\bar
q$, $F_{q\to i}$ are the transverse momentum dependent fragmentation
functions defined in \Eq{eq:TMDFF_def}, and $S_\EEC$ is the soft
function. The $\mu$ and $\nu$ are the virtuality and rapidity
renormalization scales, respectively. The dependence on rapidity scale
arises because the naive TMD fragmentation function and soft function
suffer from rapidity divergences, and need regularization and
renormalization, similar to the TMD PDF. The renormalization group evolution of each of the functions appearing in the factorization theorem will be given in \Sec{sec:RG}.

The soft function, $S_\EEC$, is defined as a vacuum matrix element of Wilson lines. Since it will play a central role in our discussion, we will carefully define $S_\EEC$, paying particular attention to the directions of the Wilson lines, and the definition of the rapidity regulator. We begin by defining four distinct soft Wilson lines
\begin{align}
S_{n+}\left( z \right) &= P \exp \left[  ig \int\limits_{-\infty}^{0} ds~ n \cdot A_{us} (z+sn) \right]\,, \\
S_{n-}^\dagger \left(z\right) &= \bar P \exp \left[  -ig \int\limits_{-\infty}^{0} ds~ n \cdot A_{us} (z+sn) \right]\,, \\
S_{n+}^\dagger \left(z\right) &= P \exp \left[  ig \int\limits^{\infty}_{0} ds~ n \cdot A_{us} (z+sn) \right]\,, \\
S_{n-} \left(z\right) &= \bar P \exp \left[  -ig \int\limits^{\infty}_{0} ds~ n \cdot A_{us} (z+sn) \right]\,.
\end{align}
Here $z^\mu$ is a reference vector defining the starting (ending) position of the Wilson line. For a detailed discussion of the Wilson line directions appearing in soft functions arising from factorization, see e.g. \cite{Chay:2004zn,Arnesen:2005nk}. The Wilson lines in different directions will be required to discuss both the soft functions appearing for the EEC, and for the color singlet $q_T$ spectrum, and to allow for an understanding of the relation between the soft functions appearing in these two cases. 

The soft function requires a rapidity regulator to be well
defined. Here we use the recently introduced rapidity regulator of
\cite{Li:2016axz}, which is implemented by displacing the Wilson lines
from the origin. This is most easily formulated in position space (impact parameter space), obtained by performing a Fourier transform in the $\perp$ momentum \cite{Collins:1981va}. Here we take $\vec b_\perp$ to be conjugate to $\vec k_\perp$. The offset of the Wilson lines is defined as
\begin{align}
y_\nu (\vec b_\perp) = ( i b_0 / \nu, i b_0/ \nu, \vec b_\perp)\,,
\end{align}
where $b_0=2e^{-\gamma_E}$. We can now define the soft function for the EEC as
\begin{align}\label{eq:EEC_soft}
S_\EEC (\vec b_\perp, \mu, \nu) =\lim_{\nu\to +\infty} \frac{1}{N_c} \mathrm{tr} \langle 0 | T \left[  S^\dagger_{\bar n +}(0) S_{n-}(0) \right] \bar T \left[ S^\dagger_{n+}\left(y_\nu(\vec b_\perp)\right) S_{\bar n -}\left(y_\nu(\vec b_\perp)\right)  \right] |0 \rangle\,. 
\end{align}
The limit appearing in \Eq{eq:EEC_soft} means to keep only the leading terms, dropping all terms that vanish as $\nu\to \infty$. The shift of the Wilson lines implements both the measurement of the $\perp$ momentum, as well as the regularization. Since the regularization can be implemented as a spacetime shift, it is well defined non-perturbatively. For a detailed discussion of the properties of this regulator, see \cite{Li:2016axz}.

The full expression for the leading power EEC cross section in the back-to-back limit can now be written as
\begin{align}
\frac{d\sigma}{dz} &= \frac{1}{2} \sum\limits_{ij} \int dx_i dx_j  x_i x_j    \int d^2 \vec k_\perp    \delta \left(  1-z -\frac{\vec k_\perp^2}{Q^2}  \right) \nn \\
& \cdot H(Q,\mu)    \int d^2 \vec k_{\perp,i}^h \int d^2 \vec k_{\perp,j}^h \int d^2 \vec k_{\perp,s}  \delta^{(2)} \left(  \vec k_\perp -\left(    \frac{\vec k_{\perp,i}^h}{x_i}+\frac{\vec k_{\perp,j}^h}{x_j}-\vec k_{\perp,s} \right)  \right) \nn \\
&\cdot F_{q\to i}(\vec k_{\perp,i}^h, x_i,\mu,\nu)  F_{q\to j}(\vec k_{\perp,j}^h, x_j,\mu,\nu)  S_\EEC(\vec k_{\perp,s},\mu,\nu) \,.
\end{align}
In its current form, this expression is still quite complicated, and furthermore, it involves the non-perturbative TMDFFs, despite the fact that the EEC is an IRC safe observable. 

To simplify this result, we can perform an operator product expansion (OPE) of the TMDFF onto the standard FF, and use the sum rule of \Eq{eq:sum_rule} to eliminate the dependence on the FF. The OPE of the TMD FF onto the standard FF is given in momentum space by \cite{Collins:2011zzd,Fleming:2003gt,Stewart:2009yx,Stewart:2010qs}
\begin{align}
F_{i \rightarrow h}(\vec k_\perp^h,z_h) =\sum \limits_j \int \frac{dz}{z^3} \cI_{ij}\left(\frac{\vec k_\perp^h}{z},\frac{z_h}{z}\right) f_{j\rightarrow h} (z, \mu) \left[  1+\cO\left( \frac{\Lambda_{\text{QCD}}^2}{(\vec k_\perp^h)^2} \right)  \right]\,.
\end{align}
Here $\cI_{ij}$ are infrared finite matching coefficients. Explicit results are given in \App{sec:tmdff-jet-function}.  Inserting this expression into \Eq{eq:J_NP}, and changing variables to
 \begin{align}
 \tau_i=\frac{x_i}{z_i}\,,  \qquad dx_i dz_i=z_i d\tau_idz_i\,.
 \end{align}
 We then find
 \begin{align}
\frac{d\sigma}{dz} &=\frac{1}{2} \sum\limits_{ij} \int d\tau_i d\tau_j  \tau_i \tau_j    \int d^2 \vec k_\perp    \delta \left(  1-z -\frac{\vec k_\perp^2}{Q^2}  \right) \nn \\
& \cdot H(Q)    \int d^2 \vec k_{\perp,i} \int d^2 \vec k_{\perp,j} \int d^2 \vec k_{\perp,s}  \delta^{(2)} \left(  \vec k_\perp -\left(    \frac{\vec k_{\perp,i}}{\tau_i}+\frac{\vec k_{\perp,j}}{\tau_j}-\vec k_{\perp,s} \right)  \right) \nn \\
&\cdot \cI_{qi}\left(\vec{k}_{\perp,i}, \tau_i\right)
  \left[ \sum\limits_h \int dz_i~ z_i~ f_{i \to h}(z_i,\mu)  \right]
  \cdot \cI_{qj}\left(\vec{k}_{\perp,j}, \tau_j\right)
  \left[ \sum\limits_{h'} \int dz_j~ z_j~ f_{j \to h'}(z_j,\mu)  \right]
\nn\\
& \cdot   S_\EEC(\vec k_{\perp,s}) \,,
\end{align}
where we have changed the convolution from hadronic transverse momentum $\vec k_{\perp,i(j)}^h$ to partonic transverse momentum $\vec k_{\perp,i(j)}$. The relation between hadronic and partonic transverse momentum is given by  $\vec k_{\perp,i}^h = z_i \vec k_{\perp,i}$ and
$\vec k_{\perp,j}^h = z_j \vec k_{\perp,j}$, which hold up to $\Ord(\Lambda_{\rm QCD})$. We also use this relation to rewrite the measurement function. It then allows us to use the momentum-conservation sum rule
 \begin{align}
   \sum\limits_h \int dz~ z~ f_{j \to h}(z,\mu) =1\,,
 \end{align}
to cancel non-perturbative fragmentation functions, and we have
\begin{align}
\frac{d\sigma}{dz} &=\frac{1}{2} \sum\limits_{ij} \int d\tau_i d\tau_j ~ \tau_i \tau_j    \int d^2 \vec k_\perp    \delta \left(  1-z -\frac{\vec k_\perp^2}{Q^2}  \right) \nn \\
& \cdot H(Q)    \int d^2 \vec k_{\perp,i} \int d^2 \vec k_{\perp,j} \int d^2 \vec k_{\perp,s}  \delta^{(2)} \left(  \vec k_\perp -\left(    \frac{\vec k_{\perp,i}}{\tau_i}+\frac{\vec k_{\perp,j}}{\tau_j}-\vec k_{\perp,s} \right)  \right) \nn \\
&\cdot \cI_{qi}(\vec{k}_{\perp,i},\tau_i)     \cdot \cI_{qj}(\vec{k}_{\perp,j},\tau_j)     S_\EEC(\vec k_{\perp,s}) \,.
\end{align}
This makes it clear that what we have is an expression in terms of the perturbative matching coefficients for the TMDFFs, $\cI_{ij}$, which are by construction IR finite. This is of course not a surprise, since the EEC observable is IRC safe, however, it is interesting to see explicitly how it arises from the sum rule for the FFs.

We can further simplify the convolution structure by transforming to impact parameter space \cite{Collins:1981va}. In addition to simplifying the convolution in the $k_{\perp,i}$ variables, as is familiar from the case of $q_T$ factorization, here we will find that this also simplifies the integrals over the momentum fractions $x_i$. Using the Fourier representation of delta function, we can write
\begin{align}
\delta^{(2)}  \left(  \vec k_\perp -\left(    \frac{\vec k_{\perp,i}}{\tau_i}+\frac{\vec k_{\perp,j}}{\tau_j}-\vec k_{\perp,s} \right) \right) 
 = \int \frac{d^2\vec{b}_\perp}{(2 \pi)^2} ~ \exp\left[-i \vec{b}_\perp \cdot \vec{k}_\perp + i \vec{b}_\perp \cdot \left(   \frac{\vec k_{\perp,i}}{\tau_i}+\frac{\vec k_{\perp,j}}{\tau_j}-\vec k_{\perp,s}\right)\right] \,.
\end{align}
The momentum convolutions are now in complete factorized form. We define the Fourier-transformed matching coefficients and soft function as
\begin{align}
\cI_{q i}(\vec b_{\perp,i}, x_i,\mu,\nu) &= \int d^2 \vec k_{\perp, i} ~\cI_{q i}(\vec k_{\perp,i}, x_i,\mu,\nu) e^{i \vec b_{\perp,i}\cdot \vec {k}_\perp,i}\,, \nn \\
S_{\EEC}(\vec b_{\perp,s},\mu,\nu) &= \int d^2 \vec k_{\perp, s}~  S_\EEC (\vec k_{\perp,s},\mu,\nu) e^{i \vec b_{\perp,s}\cdot \vec {k}_\perp,s}\,,
\end{align}
where to simplify notation, we use only the argument of the function to indicate that it is Fourier transformed. This allows us to simplify our factorized expression to 
\begin{align}
\frac{d\sigma}{dz} &=\frac{1}{2} \sum\limits_{ij} \int dx_i dx_j  x_i x_j    \int d^2 \vec k_\perp    \delta \left(  1-z -\frac{\vec k_\perp^2}{Q^2}  \right) \\
& \cdot H(Q)    \int \frac{ d^2 \vec b_\perp}{(2 \pi)^2} e^{-i \vec b_\perp \cdot \vec k_\perp}
\cdot \cI_{q i}\left(\frac{\vec b_{\perp}}{x_i}, x_i,\mu,\nu\right)  \cI_{q j} \left(\frac{\vec b_{\perp}}{x_j}, x_j,\mu,\nu\right)  S_\EEC(\vec b_{\perp},\mu,\nu) \nn\\
&=  \int d^2 \vec k_\perp    \delta \left(  1-z -\frac{\vec k_\perp^2}{Q^2}  \right)   \int \frac{d^2 \vec b_\perp}{(2 \pi)^2} e^{-i \vec b_\perp \cdot \vec k_\perp} \cdot H(Q,\mu)    \\
& 
\cdot \left[ \sum_i \int dx_i~ x_i ~\cI_{q i}\left(\frac{\vec b_{\perp}}{x_i}, x_i,\mu,\nu\right) \right]  \left[ \sum_j \int dx_j~ x_j ~ \cI_{qj} \left(\frac{\vec b_{\perp}}{x_j}, x_j,\mu,\nu\right) \right]  S_\EEC(\vec b_{\perp},\mu,\nu)\,. \nn
\end{align}
In this form, it is clear that only certain flavor summed moments of the matching coefficients for the fragmentation functions appear, and furthermore, that the integrals in the momentum fraction variables, $x_i$, are factorized. In particular, we can define the quark jet function relevant for the EEC as
\begin{align}\label{eq:J_NP}
J^q_\EEC(\vec b_\perp) = \sum\limits_i  \int \limits_0^1dx~ x~ \cI_{qi}\left(\frac{\vec b_\perp}{x},x\right)\,,
\end{align}
and similarly for the anti-quark jet function. The one-loop result for $J^q_\EEC$ is given in \App{sec:tmdff-jet-function}. This allows us to write our final factorized expression as
\begin{equation}\label{eq:fact_final}
\hspace{-0.35cm}\boxed{\frac{d\sigma}{dz}= \frac{1}{2}  \int d^2 \vec k_\perp \int \frac{d^2 \vec b_\perp}{(2 \pi)^2} e^{-i \vec b_\perp \cdot \vec k_\perp} H(Q,\mu)  J^q_\EEC (\vec b_\perp,\mu,\nu) J^{\bar q}_\EEC (\vec b_\perp,\mu,\nu) S_\EEC(\vec b_\perp,\mu,\nu)  \delta \left( 1-z- \frac{\vec k_\perp^2}{Q^2} \right )}\,.
\end{equation}
This provides a fully factorized description of the EEC in the back-to-back region into hard, jet and soft functions, and is one of the main results of this paper. We verify that this produces the known logarithmic structure at NNLO in \App{sec:NNLO_check}.

We find it interesting that this factorization theorem of \Eq{eq:fact_final} is as close as possible to a direct crossing of the factorization theorem for $q_T$ for color singlet production,\footnote{It would also be interesting to study semi-inclusive DIS with measured transverse momenta of an identified outgoing hadron. In this case, while it has been argued that the partially crossed soft function should be used \cite{Meng:1995yn,Ji:2004wu}, the analysis of \cite{Collins:1997sr} indicates that future pointing Wilson lines should be used. We leave a study of this question in our framework to future work.  We thank John Collins for discussions on this point.} which can be written in impact parameter space as
\begin{align}\label{eq:pt_fact}
\frac{1}{\sigma} \frac{d^3 \sigma}{d^2 \vec q_T dY dQ^2} = H(Q,\mu) \int \frac{d^2 \vec b_\perp}{(2\pi)^2} e^{i\vec b_\perp \cdot \vec q_T} \left[  B \times B  \right] (\vec b_\perp, \mu, \nu) S_\perp(\vec b_\perp, \mu, \nu)\,.
\end{align}
Here, instead of TMDFFs, transverse momentum dependent beam functions (also known as TMDPDFs) appear, and the soft function, referred to as the TMD soft function, is identical to the EEC soft function up to the direction of the Wilson lines in its definition. Explicitly, for the soft function, we have
\begin{align}\label{eq:qT_soft}
S_\perp (\vec b_\perp, \mu, \nu) =\lim_{\nu\to +\infty} \frac{1}{N_c} \mathrm{tr}\langle 0 | T \left[  S^\dagger_{\bar n -}(0) S_{n+}(0) \right] \bar T \left[ S^\dagger_{n-}\left(y_\nu(\vec b_\perp)\right) S_{\bar n +}\left(y_\nu(\vec b_\perp)\right)  \right] |0 \rangle\,. 
\end{align}
The precise definitions of the beam functions will not be important for the present discussion, but can be found in \cite{Gehrmann:2014yya,Stewart:2009yx,Gehrmann:2012ze}.

The key reason for the utility of this factorization theorem of \Eq{eq:fact_final} is that all the ingredients are related (or identical) to other functions that have been calculated to high perturbative accuracy, namely the TMDFFs, and the TMD soft function. This will allow us to directly use these results to improve the perturbative understanding of the EEC observable. This ability to relate different functions highlights a benefit of operator based factorization theorems.

\subsection{Renormalization Group Evolution} \label{sec:RG}

Large logarithms in the back-to-back region can be resummed by the renormalization group evolution of the functions appearing in the factorization theorem of \Eq{eq:fact_final}. Since this factorization theorem is constructed from well known objects, namely TMDFFs and the TMD soft function, we can immediately write down their renormalization group evolution. The $q_T$ dependent beam function and soft function were computed in the $\eta$ regulator of \cite{Chiu:2011qc,Chiu:2012ir} to NNLO \cite{Luebbert:2016itl}. The NNLO TMDPDF and soft function were calculated in \cite{Gehrmann:2012ze,Gehrmann:2014yya,Echevarria:2015byo}. The unpolarized TMDFF at NNLO was calculated in \cite{Echevarria:2016scs}, from which it is possible to obtain the EEC jet function using \Eq{eq:J_NP}. The $q_T$ dependent beam and jet functions will be calculated in the exponential regulator of  \cite{Li:2016axz} that was used for the calculation of the three-loop soft function for color singlet $q_T$ \cite{Li:2016ctv} in a future publication \cite{HXZ:forthcoming}.

The hard function satisfies a multiplicative RGE in $\mu$
\begin{align}
\mu \frac{d}{d\mu} H(Q,\mu) =2 \left[\gcusp(\alpha_s) \ln\frac{Q^2}{\mu^2} +  \gamma^H(\alpha_s) \right] H(Q,\mu)\,,
\end{align}
and is independent of $\nu$. Here $\Gamma_{\text{cusp}}$ is the cusp anomalous dimension \cite{Korchemsky:1987wg} (which is known analytically to three-loop order \cite{Moch:2004pa}, and numerically to four loops \cite{Moch:2017uml}), and $\gamma^H$ is the non-cusp anomalous dimension of hard function, which can be found, for example, in \cite{Gehrmann:2014yya}. The hard function is independent of the IR measurement, and its anomalous dimension can be obtained from the quark form factor, which is known to three-loops \cite{Baikov:2009bg,Lee:2010cga,Gehrmann:2010ue}. Since the result is well known (see e.g. \cite{Abbate:2010xh}), we do not explicitly give it here.

The EEC soft function satisfies RG equations in $\mu$
\begin{align}
\mu \frac{dS_\EEC(\vec b_\perp, \mu, \nu)}{d\mu} = \left[  2 \Gamma_{\text{cusp}}(\alpha_s) \ln \frac{\mu^2}{\nu^2} -2\gamma^s_\EEC (\alpha_s) \right] S_\EEC (\vec b_\perp, \mu, \nu)\,,
\end{align}
and in $\nu$~\cite{Chiu:2011qc,Chiu:2012ir,Li:2016ctv}
\begin{align}\label{eq:nu_RG_S}
\nu \frac{dS_\EEC(\vec b_\perp, \mu, \nu)}{d\nu}= 2 \left[ -  \int\limits_{b_0^2/\vec b_\perp^2}^{\mu^2} \frac{d\bar\mu^2}{\bar\mu^2} \Gamma_{\text{cusp}}(\alpha_s(\bar\mu))  +\gamma^r_\EEC(\alpha_s(b_0/|\vec b_\perp|))  \right] S_\EEC(\vec b_\perp, \mu, \nu)\,.
\end{align}
The anomalous dimensions $\gamma^s_\EEC$ and $\gamma^r_\EEC$ are known perturbatively to three-loops, and will be given in \Sec{sec:anom_dim}.

The matching coefficients, $\cI_{ik}$, for the TMDFFs satisfy the $\mu$ RG
\begin{align}
\mu \frac{d\cI_{ik}(\vec b_\perp/z, z,\mu,\nu)}{d\mu}&=  \left[- \Gamma_{\text{cusp}}(\alpha_s)  \ln \frac{Q^2}{\nu^2} +2\gamma^J_\EEC(\alpha_s)    \right] \cI_{ik}(\vec b_\perp/z, z,\mu,\nu) \\
& \hspace{4cm}-2\sum\limits_j  \cI_{ij}(\vec b_\perp/z, z,\mu, \nu) \otimes P_{jk}(z,\alpha_s)\,, \nn
\end{align}
where the convolution is defined as
\begin{align}
A(x) \otimes B(x) =\int\limits_0^1 dy \int\limits_0^1 dz \delta(x-yz) A(y) B(z)\,.
\end{align}
Note that the coefficient functions depend on the impact parameter through $\vec{b}_\perp /z$ in the argument. The additional $1/z$ factor compared with the more traditional TMDFF evolution comes from different convention in the normalization of fragmentation function.~\footnote{We thank Alexey Vladimirov for pointing out to us the standard though unusual normalization of the TMD fragmentation functions.} Additional discussion on this point will be presented in Ref.~\cite{HXZ:forthcoming}.
The $\nu$ RG is given by
\begin{align}\label{eq:nu_RG_FF}
\nu \frac{d\cI_{ik}(\vec b_\perp/z, z, \mu,\nu)}{d\nu}=  \left[  \int\limits_{b_0^2/\vec b_\perp^2}^{\mu^2} \frac{d\bar{\mu}^2}{\bar{\mu}^2} \Gamma_{\text{cusp}}(\alpha_s(\bar{\mu}))  - \gamma^r_\EEC(\alpha_s(b_0/|\vec b_\perp|))  \right] \cI_{ik}(\vec b_\perp/z,z,\mu,\nu)\,.
\end{align}
Here $P_{jk}$ are the time-like $j \to k$ splitting functions, which are known to three loops \cite{Vogt:2004mw,Moch:2004pa} and for the non-singlet case to four loops in the large $N_c$ limit~\cite{Moch:2017uml}. The anomalous dimension $\gamma^J_\EEC$ is also known to three-loops due to the consistency of the factorization, as will be discussed shortly.

Using the known RG evolution equations for the TMDFFs we can derive the RG evolution equations for the jet function $J^q_\EEC (\vec b_\perp, \mu,\nu)$ appearing in our factorization formula for the EEC. We have
\begin{align}
\label{eq:jetevolutiona}
\mu \frac{dJ^q_\EEC(\vec b_\perp, \mu,\nu)}{ d\mu} &=\sum \limits_k \int\limits_0^1 dx \, x \, \left \{    \left[- \Gamma_{\text{cusp}}(\alpha_s)  \ln \frac{Q^2}{\nu^2} +2\gamma^J_\EEC(\alpha_s)    \right] \cI_{qk}(\vec b_\perp/x, x, \mu,\nu) \right.\nn \\
&\hspace{4cm}\left. -2\sum\limits_j \cI_{qj}(\vec b_\perp/x, x,\mu, \nu) \otimes P_{jk}(x,\alpha_s)  \right \}\,.
\end{align}
The second line of Eq.~\eqref{eq:jetevolutiona} can be simplified to
\begin{align}
  \label{eq:jetsumrule}
& -2   \sum_k \int_0^1 dx \, x \, \sum_j \int dy \, dz \, \cI_{qj}(\vec b_\perp/y, y, \mu, \nu) P_{j k}(z,\alpha_s)  \delta(x - y z)
\nn\\
= & -2 \sum_j \int dy\, y \, \cI_{qj}(\vec b_\perp/y,y,\mu,\nu) \sum_k \int dz\, z \, P_{j k}(z,\alpha_s) 
\nn\\
= & \; 0 \,,
\end{align}
where in the last line we have applied the momentum conservation sum rule for the time-like splitting function,
\begin{align}
\sum\limits_j \int\limits_0^1 dx~ x~ P_{i j}(x) =0\,.
\end{align}
The $\mu$ RG for the jet function now simplifies to
\begin{align}
  \label{eq:jetRGE}
\mu \frac{dJ^q_\EEC(\vec b_\perp, \mu,\nu)}{ d\mu} &=  \left[- \Gamma_{\text{cusp}}(\alpha_s)  \ln \frac{Q^2}{\nu^2} +2\gamma^J_\EEC(\alpha_s)    \right] J^q_\EEC(\vec b_\perp,\mu,\nu) \,.  
\end{align}
The anomalous dimensions for the quark and anti-quark jet functions are identical, and therefore we will simply use the notation $\gamma^J_\EEC$ for both.

The $\nu$ RG for $J^q_\EEC$ follows trivially from the $\nu$ RG for $\cI_{ij}$ in \Eq{eq:nu_RG_FF}, since it does not involve evolution in the momentum fraction. We therefore have
\begin{align}
\nu \frac{d J^q_\EEC(\vec b_\perp,  \mu,\nu)}{d\nu}=  \left[  \int\limits_{b_0^2/\vec b_\perp^2}^{\mu^2} \frac{d\bar{\mu}^2}{\bar{\mu}^2} \Gamma_{\text{cusp}}(\alpha_s(\bar{\mu}))  - \gamma^r_\EEC(\alpha_s(b_0/|\vec b_\perp|))  \right] J^q_\EEC(\vec b_\perp,\mu,\nu)\,.
\end{align}

From the RG invariance of the total cross section, we can immediately derive several relations between the different anomalous dimensions. For the $\mu$ anomalous dimensions, we have
\begin{align}\label{eq:consist_mu}
 \frac{1}{2} \gamma^H+\gamma^J_\EEC - \frac{1}{2}\gamma^s_\EEC=0\,,
\end{align}
We have already used the consistency relations for the $\nu$ anomalous dimension in writing \Eqs{eq:nu_RG_S}{eq:nu_RG_FF}, where the same $\gamma^r_\EEC$ appears in both functions.
This implies that the hard anomalous dimension, which is known and observable independent, combined with the soft anomalous dimension, which will be given in \Sec{sec:univ}, are sufficient to determine $\gamma^J_\EEC$, and hence to completely fix the renormalization group evolution for all functions required to describe the EEC in the back-to-back region.

\section{Three-Loop Anomalous Dimensions and Soft Function} \label{sec:univ}

\begin{figure}
\begin{center}
\subfloat[]{\label{fig:wilson_flip_pt}
\includegraphics[width=7cm]{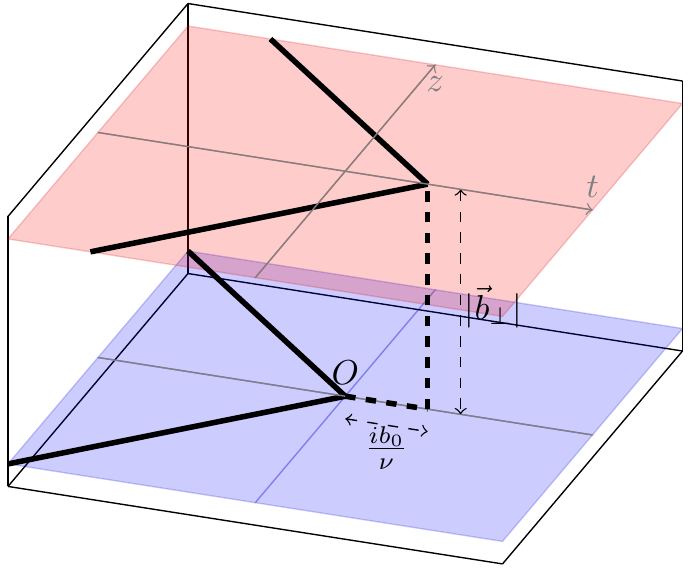}
}
\subfloat[]{\label{fig:wilson_flip_EEC}
\includegraphics[width=7cm]{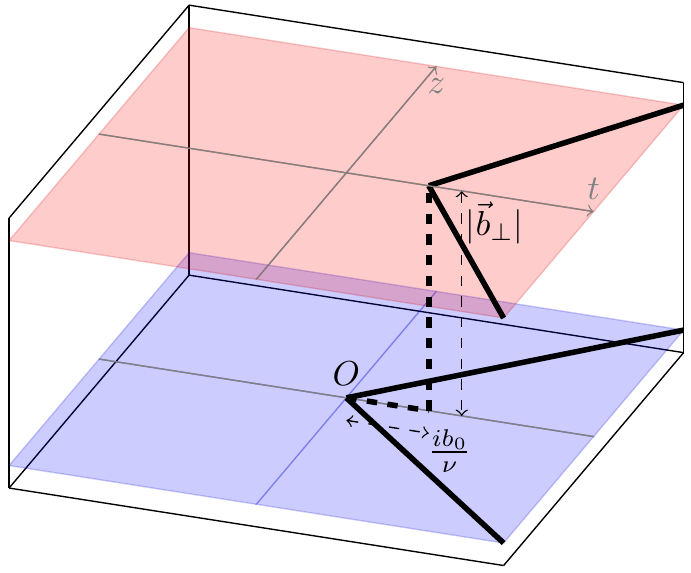}
}
\end{center}
\caption{The soft functions for $q_T$ for color singlet production in (a) and for the EEC in (b), as defined in the exponential rapidity regulator of \cite{Li:2016axz}. Solid black lines denotes Wilson lines. Both the rapidity regulator and the measurement are defined in terms of space-time shifts of the Wilson lines, allowing an all orders proof of the equivalence of these two soft functions.}
\label{fig:wilson_flip}
\end{figure}

As was noted earlier, the factorization theorem of \Eq{eq:fact_final}, which describes the singular structure of the EEC observable in the back-to-back limit, is closely related to the factorization theorem for $q_T$ for color singlet production given in \Eq{eq:pt_fact}. 
In particular, the soft functions are identical up to the directions of the Wilson lines, as illustrated in \Fig{fig:wilson_flip}. In this section, we study the relationship between the soft functions for the EEC and for $q_T$.  In \Sec{sec:anom_dim}, we use this relation to give the three-loop $\mu$ and $\nu$ anomalous dimensions for the EEC soft function, using the recently calculated results for the $q_T$ soft function. In \Sec{sec:crossing} we prove the equivalence of the soft function for $q_T$ and the EEC to all orders, i.e. the independence of the soft function on crossing the directions of the Wilson lines, which allows us to give the three-loop finite terms of the soft function for the EEC.

\subsection{Anomalous Dimensions} \label{sec:anom_dim}

The anomalous dimensions for the soft function are independent of the direction of the Wilson lines. This can be proven following \cite{Stewart:2010qs}, by noting that the jet and beam function (TMDFF and TMDPDF) anomalous dimensions can be proven to be equal from their operator definitions. The consistency relations for the anomalous dimensions in \Eq{eq:consist_mu} then imply that the anomalous dimensions of the soft function must be independent of the directions of the Wilson lines to all orders. We therefore have the relations
\begin{align}
\gamma^r_{\text{EEC}}= \gamma^r_{q_T}\equiv \gamma^r\,, \qquad   \gamma^s_{\text{EEC}}= \gamma^s_{q_T}\equiv \gamma^s\,.
\end{align}
Due to their equivalence, and to simplify our notation, we will drop the subscripts and simply write $\gamma^r$ and $\gamma^s$.
The one- and two-loop anomalous dimensions were calculated long ago \cite{Davies:1984hs,Davies:1984sp,deFlorian:2000pr}, while the three-loop anomalous dimensions were calculated quite recently.
We expand the anomalous dimensions perturbatively as
\begin{align}
\gamma^r=\sum\limits_{n\geq 0} \left(  \frac{\alpha_s}{4\pi} \right)^{n+1} \gamma^r_n\,, \qquad \gamma^s=\sum\limits_{n\geq 0}  \left(  \frac{\alpha_s}{4\pi} \right)^{n+1} \gamma^s_n\,.
\end{align}
The soft anomalous dimensions up to three-loops are~\cite{Li:2014afw}
\begin{align}
    \gsoft_{0} &=  \, 0\,,
\nbrk
  \gsoft_{1} &= \, C_A C_F \left(-\frac{808}{27}+\frac{22}{3}\zeta_2+28 \zeta_3\right)+C_F n_f \left(\frac{112}{27}-\frac{4}{3}\zeta_2\right)\,,
\nbrk
  \gsoft_{2} &=  \, C_A^2 C_F \left(-\frac{136781}{729}+\frac{12650}{81}\zeta_2+\frac{1316}{3}\zeta_3-176\zeta_4-192 \zeta_5-\frac{176}{3} \zeta_3\zeta_2\right)
\brk
+C_A C_F n_f \left(\frac{11842}{729}-\frac{2828}{81}\zeta_2-\frac{728}{27}\zeta_3+48 \zeta_4\right) \brk
+C_F^2 n_f \left(\frac{1711}{27}-4\zeta_2-\frac{304}{9}\zeta_3-16 \zeta_4\right)
+C_F n_f^2 \left(\frac{2080}{729}+\frac{40}{27}\zeta_2-\frac{112}{27}\zeta_3\right)\,.
\label{eq:c2}
\end{align}
The rapidity anomalous dimensions up to three-loops are \cite{Li:2016ctv}
\begin{align}
  \label{eq:8}
  \ugr_0 &=  \, 0\,,
\nbrk
\ugr_1 &= \, C_F C_A  \left(-\frac{808}{27}+28 \zeta_3\right)+C_F n_f\frac{112}{27}\,,
\nbrk
\ugr_2 &= \, C_F C_A^2 \left(-\frac{297029}{729} +\frac{6392}{81}\zeta_2+\frac{12328}{27} \zeta_3+\frac{154}{3}\zeta_4-192\zeta_5-\frac{176}{3} \zeta_3\zeta_2 \right) \brk
+ C_F C_A n_f \left(\frac{62626}{729}-\frac{824}{81}\zeta_2-\frac{904}{27}\zeta_3+\frac{20}{3}\zeta_4
\right) 
+ C_F n_f^2 \left(-\frac{1856}{729}-\frac{32}{9} \zeta_3 \right)
\brk
+ C_F^2 n_f \left(\frac{1711}{27} -\frac{304}{9}\zeta_3 -16 \zeta_4 \right)\,.
\end{align} 
Following the original calculation of the three-loop result in \cite{Li:2016ctv}, this result was verified in \cite{Vladimirov:2016dll} using an equivalence between rapidity and virtuality anomalous dimensions \cite{Vladimirov:2016dll,Vladimirov:2017ksc}. Using the consistency relations of \Eq{eq:consist_mu}, along with the known result for the hard anomalous dimension, this completely determines all anomalous dimensions governing the RG evolution of the EEC in the back-to-back region, and allows for resummation to N$^3$LL.

Interestingly, in planar $\cN=4$ SYM the result for the rapidity anomalous dimension can be extended to higher orders.  Using the equivalence between the rapidity anomalous dimension and the eikonal collinear anomalous dimension \cite{Vladimirov:2016dll,Vladimirov:2017ksc}, we can use the results of \cite{Dixon:2008gr} to relate $\gamma^r$ and the collinear anomalous dimension $\cG_0$ as
\begin{align}
\gamma^r= -\cG_0+2B\,.
\end{align}
Here $B$ is the virtual anomalous dimension, i.e. the coefficient of $\delta(1-x)$ in the DGLAP kernel. It is known to all orders in planar $\cN=4$ SYM using integrability \cite{Freyhult:2007pz,Freyhult:2009my,Fioravanti:2009xt}. Remarkably, the collinear anomalous dimension was recently analytically computed to four-loops \cite{Dixon:2017nat} (it was computed numerically to four-loops in \cite{Cachazo:2007ad}), providing also the rapidity anomalous dimension for the EEC at this order. The knowledge of these anomalous dimensions to such high orders, along with the hope that they may be computed to all orders using integrability, makes the EEC an interesting playground for studying the perturbative structure of resummation at high orders for a physical observable.

\subsection{Equivalence of Soft Functions and the Three-Loop Boundary Condition} \label{sec:crossing}

While the anomalous dimensions of the soft functions are independent of the direction of the Wilson lines, this is not in general true for the full soft function due to the presence of Glauber modes. At one- and two-loops, it has been shown that a wide class of soft functions are independent of the directions of the Wilson lines \cite{Kang:2015moa}. This result can also be seen using an EFT approach \cite{Rothstein:2016bsq} that allows the Glauber region to be separated. Up to two-loops, Glaubers contribute at most a phase ($i \pi$), which cancels out of the squared amplitude. However, starting at three-loops, which is the order of interest in the current paper, such Glauber contributions could begin to contribute, making the soft function depend on the direction of the Wilson lines. For a general soft function, this must be assumed to be true. To be able to achieve N$^3$LL' accuracy (where the prime denotes the inclusion of the three-loop boundary condition for the soft and collinear functions, see e.g. \cite{Almeida:2014uva} for a discussion of order counting), we would like to understand whether or not the $q_T$ and EEC soft functions are identical.\footnote{We note that the fact that   Glaubers do not contribute to the EEC or color singlet $q_T$ distributions is a distinct statement from whether or not the soft function is independent of the Wilson line directions. The statement that Glaubers cancel in a physical observable should be more precisely stated as the fact that they can be absorbed into the soft or collinear sectors by an appropriate choice of Wilson line directions. In the language of CSS, this is the statement of whether contours can be deformed out of the Glauber region, and in the EFT language it is related to whether the choice of Wilson line direction can be made such that their is a cancellation between the Glauber zero-bin and the Glaubers themselves. Since these proofs force a direction of the Wilson lines, they cannot also be used to prove independence on the direction of the Wilson lines. This would amount to circular reasoning.}

It was argued in \cite{Collins:2004nx,Collins:1992kk,Collins:2002kn} that the $q_T$ soft function is independent of the direction of the Wilson lines.\footnote{Arguments similar to those presented here, using spacetime symmetries to relate soft and collinear functions have been discussed in great detail in \cite{Collins:2004nx,Collins:1992kk,Collins:2002kn,Brodsky:2002rv,Brodsky:2002cx,Belitsky:2002sm,Boer:2003cm,Bomhof:2004aw,Bomhof:2006dp}, particularly in relation to the Sivers effect \cite{Sivers:1989cc}.} While we agree with the conclusions of \cite{Collins:2004nx,Collins:1992kk,Collins:2002kn}, details related to the time ordering and the regularization of the matrix element, which can lead to subtleties, were not made explicit.\footnote{For more detailed discussions and an alternative treatment of these issues to that presented here, see \cite{Collins:2011zzd}.} Here we will use the exponential regulator of  \cite{Li:2016axz,Li:2016ctv} to prove the independence of the soft function on crossing the Wilson line directions. However, we still believe that it would be an interesting exercise to explicitly compute the Glauber contributions in the EFT approach of \cite{Rothstein:2016bsq} to understand their invariance under the crossing of the Wilson lines. Some of the required integrals were performed in \cite{Schwartz:2017nmr}.

Our proof of the all orders equivalence of the soft function is specific to the $q_T$ soft function with Wilson lines along back-to-back directions, as well as to the particular form of the regulator of  \cite{Li:2016axz,Li:2016ctv}. Most importantly, both the measurement function, and the regulator take the form of a spacetime shift on the Wilson lines appearing in the soft function. This is shown in \Fig{fig:wilson_flip}. This is specific to the $q_T$ measurement function, and also allows the regulator to be formulated to all orders (and non-perturbatively) greatly simplifying the proof.\footnote{In particular, it is much simpler than the case of hemisphere soft functions considered in \cite{Kang:2015moa}, where the measurement function cannot in general be formulated as a spacetime shift.}

For convenience, we recall the definitions of the soft functions for $q_T$ and the EEC. For the case of the EEC, we have
\begin{align}\label{eq:soft_eec}
S_\EEC (\vec b_\perp, \mu, \nu) =\lim_{\nu\to +\infty} \frac{\mathrm{tr}}{N_c} \langle 0 | T \left[  S^\dagger_{\bar n +}(0) S_{n-}(0) \right] \bar T \left[ S^\dagger_{n+}\left(y_\nu(\vec b_\perp)\right) S_{\bar n -}\left(y_\nu(\vec b_\perp)\right)  \right] |0 \rangle\,,
\end{align}
and for the case of the color-singlet $q_T$ soft function
\begin{align}\label{eq:soft_qt}
S_\perp (\vec b_\perp, \mu, \nu) =\lim_{\nu\to +\infty} \frac{\mathrm{tr}}{N_c} \langle 0 | T \left[  S^\dagger_{\bar n -}(0) S_{n+}(0) \right] \bar T \left[ S^\dagger_{n-}\left(y_\nu(\vec b_\perp)\right) S_{\bar n +}\left(y_\nu(\vec b_\perp)\right)  \right] |0 \rangle\,. 
\end{align}
Again, we emphasize that due to the particular nature of the measurement, and the implementation of the regulator as a spacetime shift, this is a vacuum matrix element of (shifted) Wilson lines. Here we have also made the (anti-) time ordering explicit (For a discussion of the importance of the time-ordering, see  \cite{Stewart:2009yx}). The time ordering must be treated carefully, since when using time reversal arguments to flip the directions of the Wilson lines, the time ordering also flips, as can be seen for a simple bosonic field
\begin{align}
T[\phi(t_1) \phi(t_2)] &=\phi(t_1) \phi(t_2) \Theta(t_1 -t_2) +\phi(t_2) \phi(t_1) \Theta(t_2-t_1)\nn \\
&\xrightarrow{\text{T}}  \phi(-t_1) \phi(-t_2) \Theta(-t_1 +t_2) +\phi(-t_2) \phi(-t_1) \Theta(-t_2+t_1)\nn \\
&=\bar T [\phi(-t_1), \phi(-t_2)]\,.
\end{align}
For general soft functions, where the regulator and measurement cannot be formulated as a shift, one has a time ordered matrix element squared, and the time ordering can disrupt the proof of Wilson line direction independence, as noted in \cite{Kang:2015moa}. However, as seen in \Eqs{eq:soft_eec}{eq:soft_qt}, for our particular soft function of interest both time ordered and anti-time ordered contributions appear in the matrix element, which will exchange under time reversal.

To prove the universality of the soft function we start with the EEC soft function, and apply time reversal symmetry,\footnote{For a detailed discussion of the transformation properties of Wilson lines, see e.g. \cite{Collins:2011zzd}.} using the fact that the vacuum states are invariant
\begin{align}
S_\EEC (\vec b_\perp, \mu, \nu) &=\lim_{\nu\to +\infty} \frac{\mathrm{tr}}{N_c} \langle 0 | T \left[  S^\dagger_{\bar n +}(0) S_{n-}(0) \right] \bar T \left[ S^\dagger_{n+}\left(y_\nu(\vec b_\perp)\right) S_{\bar n -}\left(y_\nu(\vec b_\perp)\right)  \right] |0 \rangle\nn \\
&\xrightarrow{\text{T}} \lim_{\nu\to +\infty} \frac{\mathrm{tr}}{N_c} \langle 0 |  T \left[ S_{n+}\left(y^T_\nu(\vec b_\perp)\right) S_{\bar n -}^\dagger\left(y^T_\nu(\vec b_\perp)\right)  \right]  \bar T \left[  S_{\bar n +}(0) S_{n-}^\dagger(0) \right] |0 \rangle\,.
\end{align}
Here the time reversal changes the displacement of the Wilson lines\footnote{We have used the superscript notation for the transformed vector to distinguish it from the time ordering operator.}
\begin{align}
y_\nu (\vec b_\perp) &= ( i b_0 / \nu, i b_0/ \nu, \vec b_\perp)\nn \\
&\xrightarrow{\text{T}} y^T_\nu (\vec b_\perp)\equiv  (- i b_0 / \nu, -i b_0/ \nu, \vec b_\perp)\,.
\end{align}
We can now use the translation invariance of the matrix element, combined with the fact that the soft function depends only on $\vec b_\perp^2$, to translate the arguments back to the original positions defining the $S_{\perp}$ soft function. This set of transformations can also easily be understood by looking at the positions of the two Wilson lines in \Fig{fig:wilson_flip}. We therefore obtain
\begin{align}
S_\EEC (\vec b_\perp, \mu, \nu)=S_\perp (\vec b_\perp, \mu, \nu)\,.
\end{align}
In summary the proof used that the measurement and regulator were formulated as spacetime shifts, that there were only two Wilson line directions, that the soft function is independent of $n \leftrightarrow \bar n$,  depends only on $x=-\vec b_\perp^2 \nu^2/b_0^2$, and that one has time reversal invariance, and translation invariance. It is therefore quite specific to the particular case of interest, and we do not make claims for more general soft functions.

Using this equivalence, and the recently computed three-loop result for the $q_T$ soft function \cite{Li:2016ctv}, we can give the three-loop constant for the EEC soft function, which acts as the boundary condition for the RG evolution. Using the non-Abelian exponentiation theorem \cite{Sterman:1981jc,Gatheral:1983cz,Frenkel:1984pz}, which is preserved by the exponential regulator, we can write the soft function as
\begin{align}
S_\EEC(\vec b_\perp, \mu, \nu)= \exp \left[  \left( \frac{\alpha_s}{4\pi} \right) S^\EEC_1+ \left( \frac{\alpha_s}{4\pi} \right)^2 S^\EEC_2 + \left( \frac{\alpha_s}{4\pi} \right)^3 S^\EEC_3 +\cO(\alpha_s^4)  \right]\,.
\end{align}
The boundary conditions are then given by the soft function evaluated at its natural scales
\begin{align}
c_i^\EEC \equiv S^\EEC_i \left (\vec b_\perp, \mu=\frac{b_0}{|\vec b_\perp|},\nu=\frac{b_0}{|\vec b_\perp|}   \right)\,.
\end{align}
Using the results of \cite{Li:2016ctv}, we can now give the explicit result for the EEC soft function constant to three-loops:
\begin{align}
\label{eq:9}
c^{\EEC}_1 &= -2C_F \zeta_2 \,, \nn \\
c^{\EEC}_2 &= C_A C_F \left(\frac{2428}{81}  -\frac{67}{3}\zeta_2 -\frac{154}{9} \zeta_3 +10 \zeta_4 \right)\, \nn \\
&+C_F n_f \left(  -\frac{328}{81} +\frac{10}{3}\zeta_2 +\frac{28}{9} \zeta_3    \right)\,, \nn \\
c^{\EEC}_3 &= \, C_F C_A^2 \left(\frac{5211949}{13122}-\frac{297481}{729} \zeta_2-\frac{151132}{243} \zeta_3+\frac{3649 }{27}\zeta_4 \right. \nn\\
&\hspace{7.5cm}\left.+\frac{1804}{9}\zeta_5+\frac{1100}{9} \zeta_2 \zeta_3-\frac{3086 }{27}\zeta_6
+ \frac{928}{9} \zeta_3^2\right)
\brk
+ C_F C_A n_f \left(-\frac{412765}{6561}+\frac{74530}{729}\zeta_2+\frac{8152}{81}\zeta_3-\frac{416}{27}\zeta_4
-\frac{184}{3}\zeta_5+\frac{40}{9} \zeta_3 \zeta_2 \right)
\brk
+ C_F^2 n_f \left(-\frac{42727}{486} +\frac{275}{9}\zeta_2+\frac{3488}{81}\zeta_3+\frac{152}{9}\zeta_4+\frac{224}{9}\zeta_5-\frac{80}{3} \zeta_3 \zeta_2
\right)
\brk
+ C_F n_f^2 \left( -\frac{256}{6561}-\frac{136}{27} \zeta_2-\frac{560}{243}\zeta_3-\frac{44}{27}\zeta_4 \right)\,.
\end{align}
The full result for the soft function evaluated at a general scale is given in \App{sec:NNLO_check}.
This is an important ingredient for resummation to N$^3$LL' accuracy for the EEC. Here the superscript $'$ denotes the inclusion of the constant terms in the functions in the EFT. It is often found that this improves the matching to the NNLO fixed order result, see e.g. \cite{Abbate:2010xh,Hoang:2014wka}.  This represents the state of the art for any event shape observable in QCD, and furthermore, this is the first time that this accuracy has been achieved for a recoil sensitive ($\SCETii$) $e^+e^-$ event shape. 

\section{Resummation Formula} \label{sec:resum}

The factorization theorem in Eq.~\eqref{eq:fact_final} can be used to
resum all large logarithms of $1-z$ appearing in the back-to-back region through RG evolution in both rapidity and virtuality. By rotational symmetry,  we can integrate out
$\vec k_\perp$ and the angular component of $\vec b_\perp$, giving
\begin{align}
  \label{eq:integrateoutk}
  \frac{d\sigma}{dz} =\frac{1}{2} \int\limits_0^\infty  \frac{b \, db}{2}  \, J_0(b Q  \sqrt{1-z}) H(Q,\mu) j^q_\EEC (b,\mu,\nu) j^{\bar q}_\EEC
  (b, \mu,\nu) S_\EEC( b,\mu,\nu) \,,
\end{align}
where $J_0(x)$ is the Bessel function of the first kind, and we have made it clear that the jet and soft functions only
depend on the magnitude of $\vec b_\perp$, $b=\sqrt{\vec
  b_\perp^2}$. 
  
Resummation can be achieved by setting the
renormalization and rapidity separately for each of the factorized
ingredient to minimize the large logarithms, and then evolving all
scales to a common value. The natural scales for the hard, jet and soft functions
are
\begin{align}
  \label{eq:scales}
  \mu_h = Q, \quad \mu_j = b_0/b, \quad \mu_s = b_0/b, \quad \nu_j = Q,
  \quad \nu_s = b_0/b \,.
\end{align}
Below we choose to evolve the hard function and soft function to the jet function scales.  Other choices could also be used, as guaranteed by the consistency of the anomalous dimensions. The evolution for the hard function is
\begin{align}
  \label{eq:hardevolution}
  H(Q,\mu) = H(Q,\mu_h) \exp\left[ \int\limits_{\mu_h^2}^{\mu^2}
  \frac{d \bar{\mu}^2}{\bar{\mu}^2} \left( \gcusp(\alpha_s(\bar{\mu}))
  \ln \frac{Q^2}{\bar{\mu}^2} + \gamma_H(\alpha_s(\bar{\mu}) \right) \right]\,.
\end{align}
For the soft function, we have evolution both in renormalization scale
and rapidity scale,
\begin{align}
  \label{eq:softevolution}
  S_\EEC(b,\mu,\nu) = & \;S_{\EEC}(b,\mu_s,\nu_s)
  \exp\left[\int\limits_{\mu_s^2}^{\mu^2}
  \frac{d\bar{\mu}^2}{\bar{\mu}^2} \left(\gcusp(\alpha_s(\bar \mu)) \ln
  \frac{b^2 \bar{\mu}^2}{b_0^2} - \gamma^s_\EEC(\alpha_s(\bar \mu))\right)
\right.
\nn\\
&\;\left. + \ln \frac{\nu^2}{\nu_s^2} \left( -
  \int\limits_{b_0^2/b^2}^{\mu^2} \frac{d\bar{\mu}^2}{\bar{\mu}^2}
  \gcusp (\alpha_s(\bar \mu)) + \gamma^r_\EEC (\alpha(b_0/b))\right)
\right]\,.
\end{align}
Substituting Eqs.~\eqref{eq:hardevolution} and \eqref{eq:softevolution}
into Eq.~\eqref{eq:integrateoutk}, and setting $\mu = b_0/b$, $\nu =
Q$, we obtain
\begin{align}
  \label{eq:resformula}
\frac{d\sigma}{dz} = &\; \frac{1}{4} \int\limits_0^\infty db\, b
  J_0(bQ\sqrt{1-z})H(Q,\mu_h) j^q_\EEC(b,b_0/b,Q) j^{\bar
  q}_\EEC(b,b_0/b,Q) S_\EEC( b,\mu_s, \nu_s) 
\nn\\
&\; \cdot
  \left(\frac{Q^2}{\nu_s^2}\right)^{\gamma^r_\EEC(\alpha_s(b_0/b))}
  \exp \left[ \int\limits_{\mu_s^2}^{\mu_h^2}
  \frac{d\bar{\mu}^2}{\bar{\mu}^2} \gcusp(\alpha_s(\bar \mu)) \ln
  \frac{b^2\bar{\mu}^2}{b_0^2} \right.
\nn\\
&\;
\left. +
  \int\limits_{\mu_h^2}^{b_0^2/b^2}\frac{d\bar{\mu}^2}{\bar{\mu}^2}
  \left(\gcusp(\alpha_s(\bar \mu)) \ln\frac{b^2 Q^2}{b_0^2} +
   \gamma^H (\alpha_s(\bar \mu)) \right) -
  \int\limits_{\mu_s^2}^{b_0^2/b^2}\frac{d\bar{\mu}^2}{\bar{\mu}^2}
   \gamma^s_\EEC (\alpha_s(\bar \mu))  \right]\,.
\end{align}
Eq.~\eqref{eq:resformula} gives our final formula for the resummation of large logarithms of
$1-z$ for the EEC in the back-to-back region, and is another main result of this paper. It shows the all orders resummation of logarithms of $1-z$, and we have given field theoretic definitions for all ingredients appearing in the formula, in particular, for the anomalous dimensions $\gamma_{\EEC}^r$, $\gamma^s_\EEC$, and $\gamma^H$, which control the renormalization group evolution. At each perturbative order, remaining scale uncertainties are estimated by varying $\mu_h$, $\mu_s$, and $\nu_s$ around their nominal values.

Here we have performed the resummation directly in impact parameter space. There has been recent work on understanding the resummation of $q_T$ sensitive observables in momentum space \cite{Monni:2016ktx,Ebert:2016gcn,Kang:2017cjk}. This has been done in \cite{Monni:2016ktx} using a coherent branching type formalism \cite{Banfi:2004yd,Banfi:2014sua}, and in \cite{Ebert:2016gcn} by  solving distributional evolution equations. We hope that the particularly simple form of the resummation for the EEC, and the fact that it is a non-perturbatively well defined observable even in a conformal theory, may allow it to be a useful observable for studying many of these issues.

Finally, we note that we have considered in this section only the perturbative distribution. Non-perturbative corrections to the EEC have been studied in \cite{Dokshitzer:1999sh,Fiore:1992sa,deFlorian:2004mp,Tulipant:2017ybb}. An important aspect of our factorization theorem is the operator definitions of the jet and soft functions that describe the dynamics of the EEC in the back-to-back limit. This enables non-perturbative effects to be studied, and related to other observables, in particular, $q_T$. Conversely, there has been significant interest in the non-perturbative functions appearing in the description of the $q_T$ distribution, such as $g_K(b_T)$ (see e.g. \cite{Collins:2017oxh} for definitions and a recent discussion), which is closely related to our rapidity anomalous dimenion $\gamma^r_{\EEC}$. The fact that these functions also appear in the EEC, which is an inclusive event shape, may facilitate their study.

\section{Conclusions} \label{sec:conclusions}

In this paper we have presented an analytic result for the three-loop soft function for the EEC observable in the back-to-back region. This result was derived from a new factorization theorem describing the leading power dynamics in the back-to-back region, whose soft function is identical to the case of $q_T$ for color singlet production up to the direction of the Wilson lines. This factorization theorem provides an operator level correspondence between the EEC observable, and $q_T$, which is the most important advantage of our approach compared to approaches taken previously in the literature, for example in Ref.~\cite{deFlorian:2004mp}. In Ref.~\cite{deFlorian:2004mp}, the NNLL resummation formula is established by matching a CSS like formula with the single logarithmic term at $\cO(\alpha_s^2)$ from an explicit two-loop perturbative calculation. In our formula, predictions at NNLL accuracy are fully determined using one-loop matching calculation for the soft and jet function, and the  well-known anomalous dimensions from $q_T$ resummation, thanks to the correspondence between Drell-Yan and $e^+e^-$ process as was explained in Sec.~\ref{sec:univ}. Furthermore, our formula can also predict the coefficient of $\delta(1-z)$. An explicit example at NLO is given in the \App{sec:NNLO_check}. Our factorization theorem thus enables the resummation of all large logarithms appearing in the perturbative expansion of the EEC in the back-to-back region beyond NNLL, and we provided analytic results for all anomalous dimensions to three-loop order, allowing resummation to N$^3$LL. 

The EEC is now the $q_T$ sensitive ($\SCETii$) observable about which the highest order perturbative information is known, making it a prime candidate for precision extractions of $\alpha_s$ from LEP data, which will complement those from $\SCETi$ observables. This has already been pursued recently in Ref.~\cite{Tulipant:2017ybb} at NNLL matched to NNLO, and it would be interesting to improve the perturbative precision to N$^3$LL. In addition to the anomalous dimensions presented here, the full calculation at N$^3$LL+NNLO will also require the calculation of the NNLO jet functions. This can be accomplished by crossing ingredients used in the calculation of the transverse momentum dependent beam functions, and results with the  exponential regulator used here will be presented in a future publication \cite{HXZ:forthcoming}.  Along similar lines, the distinct perturbative and non-perturbative structure as compared with recoil free observables will make the comparison of precision calculations for the EEC with Monte Carlo parton shower programs useful for improving the modeling of quark and gluon jets, as was considered for thrust in \cite{Mo:2017gzp}. 

The exceptional perturbative control of both the EEC and color singlet $q_T$ spectrum motivates an improved understanding of non-perturbative effects for $q_T$ sensitive observables. While non-perturbative effects have been studied for broadening \cite{Becher:2013iya}, $q_T$ \cite{Davies:1984sp,Davies:1984hs,Arnold:1990yk,Ladinsky:1993zn,Becher:2013iya}, groomed fragmentation \cite{Makris:2017arq}, semi-inclusive DIS \cite{Boglione:2016bph}, and the EEC \cite{Dokshitzer:1999sh,Fiore:1992sa,deFlorian:2004mp,Tulipant:2017ybb}, it has been found in a variety of studies that the standard shape function parametrizations used were not sufficient to describe non-perturbative effects \cite{deFlorian:2004mp,Tulipant:2017ybb}. It will be essential to achieve an improved understanding for precision extractions of $\alpha_s$, and we hope that this will also help in understanding the non-perturbative corrections for the $q_T$ spectrum.

There are a number of additional directions that will be interesting to pursue involving the EEC. In the $\chi\to 0$ limit, the EEC can be calculated at LL accuracy using the jet calculus \cite{Konishi:1979cb} (see also \cite{Belitsky:2013ofa}), however, it would also be interesting to formulate an operator based factorization theorem in terms of jet and soft functions, to allow improved perturbative control in this limit.  It may also be interesting to study higher point energy-energy correlations in $e^+e^-$. This has been done successfully in jet substructure \cite{Larkoski:2013eya,Larkoski:2014gra,Larkoski:2014zma,Larkoski:2015kga,Moult:2016cvt,Komiske:2017aww}, but could hopefully be done in a manner which preserves the simple perturbative structure of the EEC. Finally, the simplicity of the EEC observable may also prove useful for the study of the analytic structure of fixed order corrections to perturbative event shapes,  and of their perturbative power corrections \cite{Freedman:2013vya,Moult:2016fqy,Boughezal:2016zws,Balitsky:2017flc,Moult:2017jsg,Balitsky:2017gis}. We hope that these many interesting directions can generate renewed interest in the EEC observable.

\begin{acknowledgments}
	
We thank Lance Dixon for bringing the EEC observable to our attention, as well as for discussions regarding his $\cN=4$ results for the collinear anomalous dimension, and Duff Neill for helpful discussions and references regarding the universality of the soft function and the role of Glaubers. We thank Tong-Zhi Yang for checking some of the equations in this paper. 
We thank Alexey Vladimirov, John Collins, Lance Dixon, Anjie Gao and Markus Ebert for helpful comments on the first draft, as well as for catching a number of typos.
I.M thanks Zhejiang University for hospitality while portions of this work were performed. I.M was supported in part by the Office of High Energy Physics of the U.S. Department of Energy under Contract No. DE-AC02-05CH11231, and the LDRD Program of LBNL. H.X.Z. was supported in part by the Thousand Youth Program of China under contract 588020-X01702/076.

\end{acknowledgments}

\appendix

\section{Matching Coefficients for the TMDFF and EEC Jet Function}
\label{sec:tmdff-jet-function}

The matching coefficient $\cI_{ji}(x, \vec k_\perp)$ from the TMDFF to the conventional fragmentation
function can be calculated perturbatively as the probability of
finding a parton $i$ from a parton $j$, with momentum fraction
$x$ and transverse momentum $\vec k_\perp$ relative to the partonic
jet axis, which is aligned with the total jet three momentum. At LO the
matching coefficients are trivial,
\begin{align}
  \label{eq:loi}
  \cI_{ji}(x,\vec b_\perp) = \, \left\{
\begin{array}{ll}
\delta(1-x) & \qquad \text{if } i = j  \,,
 \\
0 & \qquad \text{if } i \neq j  \,. 
\end{array}
\right.
\end{align}
At NLO, the matching coefficients before zero-bin subtraction \cite{Manohar:2006nz} can be calculated from the LO
splitting kernel $P_{\widetilde{ik}}\to p_i k$, where $p_i$ and $k$ are
on-shell momentum. Explicitly, we have
\begin{align}
\label{eq:nloi}
  \frac{\alpha_s}{4 \pi} \cI_{ji}^{(1),\rm bare} = \, & \frac{1}{z}  
 \mu^{2\e}\lim_{\tau \to 0} \int \frac{\df^{4-2 \e} k}{(2 \pi)^{3-2 \e}} \Theta(k^0)
\delta(k^2) \delta\left( \frac{k^-}{Q} - (1-z)\right) g_s^2 \frac{1}{s_{ik}} p_{ji}^{(0)}
                                                        (z,\e) 
\nn\\
& \cdot \exp \left[ - \frac{b_0 \tau}{2} (k^+ + k^-) + i \vec b_\perp
  \cdot \vec k_\perp \right] \,,
\end{align}
where $\tau = 1/\nu$, $b_0 = 2 e^{-\gamma_E}$, and $Q = P_{\widetilde{ik}}^-$ is
the label momentum of the jet. The $1/z$ factor comes from
phase space factorization, and $s_{ik} = (p_i + k)^2 = \vec k_\perp^2/(z(1-z))$.  For the quark FF, the relevant splitting kernel are
\begin{align}
  \label{eq:16}
  p_{qq}^{(0)}(z,\e) = \, & 2 C_F \left[ \frac{1 + z^2}{1-z} - \e
  (1-z)\right] \,,
\nn\\
p_{qg}^{(0)}(z,\e) = \, & p_{qq}(1-z, \e) \,.
\end{align}
The integral in Eq.~\eqref{eq:nloi} can be done analytically in the
limit of $\tau \to 0$, keeping only the leading power terms in
$\tau$. The results are
\begin{align}
  \label{eq:17}
  \widetilde{\cI}_{qq}^{(1)} = \, & C_F \left( - 2 \Lp^2 -2 \Lp L_Q - 4 \Lp
  L_\nu + 3 \Lp - \frac{\pi^2}{3} \right) \delta(1-z)  - C_F \Lp
  P_{qq}^\zero (z) + 2 C_F ( 1- z) \,,
\nn\\
  \widetilde{\cI}_{qg}^{(1)} = \, & C_F \left( - L_\perp P_{qg}^{(0)}(z) + 2 z\right) \,,
\end{align}
where
\begin{align}
  \label{eq:18}
  P_{qq}^\zero(z) = \, & 3 \delta(1-z) + 2 \frac{1+z^2}{[1-z]_+}  \,, \nn\\
P_{qg}^\zero(z) = \, &  \frac{4 - 4 z + 2 z^2}{z} \,.
\end{align}
Note that there is no need to regularize $P_{qg}^\zero(z)$ in
Eq.~\eqref{eq:18}, since in the jet function it is weighted by $z$ in
the numerator. In Eq.~\eqref{eq:17} we have defined
\begin{align}
  \label{eq:19}
  L_\perp = \ln \frac{\vec b_\perp^2 \mu^2}{b_0^2} \,, \quad
L_\nu = \ln \frac{\nu^2}{\mu^2} \,, \quad
L_Q = \ln \frac{Q^2}{\nu^2} \,.
\end{align}
The results in Eq.~\eqref{eq:17} have a non-trivial zero-bin. In the
exponential regularization scheme~\cite{Li:2016axz}, the zero-bin is the same
as the soft function. The zero-bin can be straightforwardly removed by
dividing the fragmentation function by the soft function. We find that
the zero-bin subtracted TMDFF coefficients are
\begin{align}
  \label{eq:20}
  \cI_{qq}^\one = \, & C_F \left( -2 L_\perp L_Q + 3 L_\perp)
                       \delta(1-z) - L_\perp P_{qq}^\zero (z) + 2
                       (1-z) \right) \,, \nn\\
\cI_{qg}^\one = \, & \widetilde{\cI}_{qg}^{(1)}=C_F \left( - L_\perp P_{qg}^{(0)}(z) + 2 z\right)   \,.
\end{align}

Using these results we can compute the tree level and one-loop result for the jet function appearing in the EEC factorization theorem. Recall that it was defined as
\begin{align}\label{eq:J_NP_app}
J^q_\EEC(\vec b_\perp) = \sum\limits_i  \int \limits_0^1dx~ x~ \cI_{qi}\left(\frac{\vec b_\perp}{x},x\right)\,.
\end{align}
Using \Eq{eq:loi}, we find
\begin{align}
J^{q(0)}_\EEC(\vec b_\perp) =1\,.
\end{align}
At NLO, we can write the logarithm appearing in the splitting functions as
\begin{align}
\ln \left(  \frac{\vec b_\perp^2 \mu^2}{x^2 b_0^2}  \right) =\ln \left(  \frac{\vec b_\perp^2 \mu^2}{b_0^2} \right) -\ln \left(x^2 \right)\,.
\end{align}
The calculation of logarithmically enhanced terms is then made trivial using the sum rule for the tree level splitting functions
\begin{align}
\sum\limits_i \int\limits_0^1 dx~ x~ P_{ij}^{(0)}(x) =0\,.
\end{align}
However, the splitting functions enter the calculation of the constant, and we find
\begin{align}
c_1^J&=\int\limits_0^1 dx~ x~ C_F \left[  \ln\left(x^2\right) P^{(0)}_{qq}(x) +2(1-x)  \right] 
+\int\limits_0^1 dx~ x~ C_F \left[ \ln\left(x^2\right) P^{(0)}_{qg}(x) +2x  \right] \nn \\
&=(4-8\zeta_2) C_F\,.
\end{align}
We therefore find that the one-loop jet function for the EEC is given by
\begin{align}
J^{q(1)}_\EEC(\vec b_\perp) = C_F \left(   -2L_\perp L_Q +3 L_\perp  \right) +c_1^J\,.
\end{align}

\section{Logarithmic Structure to NNLO}
\label{sec:NNLO_check}

In this appendix we perform a check of our factorization formula for the EEC observable by reproducing the known logarithmic structure up to NNLO. We begin by collecting a number of ingredients that will be required, namely the hard, jet and soft functions, their associated anomalous dimensions, and results for vector plus functions that will allow us to treat the integrals appearing in the factorization theorem.

The full scale dependent soft function is given by
\begin{align}
  S_\EEC(\vec b_\perp,\mu,\nu) = \, & \exp\Bigg\{
\left( \frac{\alpha_s}{4\pi} \right) \Bigg[ 
c^{\EEC}_1+\frac{1}{2} \gcusp_0 L_\perp^2+\ugr_0 L_r - L_\perp \left(\gsoft_0+\gcusp_0 L_r\right)
\Bigg] 
\brk
+
\left(\frac{\alpha_s}{4\pi}\right)^2 \Bigg[
c^{\EEC}_2+\ugr_1 L_r+\frac{1}{6} \gcusp_0 L_\perp^3 \beta _0+L_\perp^2
                                \left(\frac{\gcusp_1}{2}-\frac{\gsoft_0
                                \beta _0}{2}-\frac{1}{2} \gcusp_0 L_r
                                \beta _0\right)
\brk
+L_\perp \left(-\gsoft_1+c^{\perp}_1 \beta _0+L_r \left(-\gcusp_1+\ugr_0 \beta _0\right)\right)
\Bigg]
\brk
+
\left(\frac{\alpha_s}{4\pi}\right)^3 \Bigg[
c^{\EEC}_3+\ugr_2 L_r+\frac{1}{12} \gcusp_0 L_\perp^4 \beta _0^2
\brk
                                +L_\perp^3 \left(\frac{\gcusp_1 \beta
                                _0}{3}+\frac{1}{3} -\gsoft_0 \beta
                                _0^2-\frac{1}{3} \gcusp_0 L_r \beta
                                _0^2+\frac{\gcusp_0 \beta
                                _1}{6}\right)
\brk
+L_\perp^2 \left(\frac{\gcusp_2}{2}-\gsoft_1 \beta _0+c^{\perp}_1 \beta _0^2-\frac{\gsoft_0 \beta _1}{2}+L_r \left(-\gcusp_1 \beta _0+\ugr_0 \beta _0^2-\frac{\gcusp_0 \beta _1}{2}\right)\right)
\brk
+L_\perp \left(-\gsoft_2+2 c^{\perp}_2 \beta _0+c^{\perp}_1 \beta _1+L_r
                                \left(-\gcusp_2+2 \ugr_1 \beta
                                _0+\ugr_0 \beta _1\right)\right)
\Bigg] 
+ \Ord(\alpha_s^4) \Bigg\}\,,
\end{align}
where $L_r = \ln \big(\nu^2 \vec b_\perp^{\, 2}/b^2_0\big)$ is the rapidity
logarithm, and $L_\perp = \ln (\vec b_\perp^{\, 2}\mu^2/b^2_0)$, as before. Expanded to $\cO(\alpha_s^2)$, as is relevant for our check to NNLO, we find
\begin{align}
S_\EEC(\vec b_\perp,\mu,\nu) &=1+\left( \frac{\alpha_s}{4\pi} \right) \Bigg[ 
c^{\EEC}_1+\frac{1}{2} \gcusp_0 L_\perp^2+\ugr_0 L_r+L_\perp \left(-\gsoft_0-\gcusp_0 L_r\right) 
\Bigg] \nn \\
&+
\left(\frac{\alpha_s}{4\pi}\right)^2 \Bigg[
c^{\EEC}_2+\ugr_1 L_r+\frac{1}{6} \gcusp_0 L_\perp^3 \beta _0+L_\perp^2
                                \left(\frac{\gcusp_1}{2}-\frac{\gsoft_0
                                \beta _0}{2}-\frac{1}{2} \gcusp_0 L_r
                                \beta _0\right)
\brk
+L_\perp \left(-\gsoft_1+c^{\EEC}_1 \beta _0+L_r \left(-\gcusp_1+\ugr_0 \beta _0\right)\right)
\Bigg]\nn \\
&+\frac{1}{2}\left( \frac{\alpha_s}{4\pi} \right)^2 \Bigg[ 
c^{\EEC}_1+\frac{1}{2} \gcusp_0 L_\perp^2+\ugr_0 L_r-L_\perp \left(\gsoft_0+\gcusp_0 L_r\right) 
\Bigg]^2 +\cO(\alpha_s^3)\,.
\end{align}
We will normalize the hard function so that its tree level value is $1$, by pulling out the tree level total cross section
\begin{align}
\sigma_0=\frac{4\pi \alpha^2}{Q^2} \sum_q \sigma_q e_q^2\,.
\end{align}
The scale dependent hard function is then given to two-loops by (see e.g. \cite{Becher:2008cf})
\begin{align}
H&=1+ \left(\frac{\alpha_s}{4\pi} \right)  \left(  -\frac{1}{2}\gcusp_0 L_H^2 +\gamma_0^H L_H +c_1^H  \right)  \\
&+\left(\frac{\alpha_s}{4\pi} \right)^2 \left[  \frac{1}{8} (\gcusp_0)^2 L_H^4 -\left( \frac{\beta_0 \gcusp_0}{6} +\frac{\gamma_0^H \gcusp_0}{2}  \right) L_H^3  \right. \nn \\
&\left. +\left(  \frac{(\gamma_0^H)^2}{2} +\frac{\beta_0 \gamma_0^H}{2} -\frac{\gcusp_1}{2}\right) L_H^2 +\gamma_1^H L_H +c_1^H \left( -\frac{\gcusp_0}{2}L_H^2 +L_H(\beta_0+\gamma_0^H)   \right) +c_2^H \right ] \nn \\
&+\cO(\alpha_s^3)\,, \nn
\end{align}
which is sufficient for our purposes. Here $L_H=\ln\frac{\mu^2}{Q^2}$, and the hard function constants are given by
\begin{align}
  c^{H}_1 = & \, C_F (14 \zeta_2-16)\,, \\
  c^{H}_2 = & \, C_A C_F \left(\frac{1061 \zeta_2}{9}+\frac{626
      \zeta_3}{9}-16 \zeta_4-\frac{51157}{324}\right)
+C_F^2 \left(-166 \zeta_2-60 \zeta_3+201 \zeta_4+\frac{511}{4}\right) \nn \\
&+C_F n_f \left(-\frac{182 \zeta_2}{9}+\frac{4 \zeta_3}{9}+\frac{4085}{162}\right)\,.
\end{align}
The hard anomalous dimensions are given by
\begin{align}
\gamma_0^H&=-6C_F\,, \\
\gamma_1^H&=C_F^2(-3+24\zeta_2-48 \zeta_3) +C_F C_A \left(  -\frac{961}{27} -22\zeta_2+52\zeta_3 \right)+ C_F T_F n_f \left( \frac{260}{27}+8\zeta_2  \right)\,. 
\end{align}
We will also need the one-loop running of $\alpha_s$, which is given by
\begin{align}
\alpha_s(\mu)=\alpha_s(\mu_R) \left( 1-\frac{\alpha_s(\mu_R)}{4\pi} \beta_0  \ln\left( \frac{\mu^2}{\mu_R^2}\right) +\cO(\alpha_s^2)  \right)\,,
\end{align}
where
\begin{align}
\beta_0=\frac{11}{3}C_A -\frac{2}{3}n_f\,.
\end{align}
The quark cusp anomalous dimensions are \cite{Korchemsky:1987wg}
\begin{align}
\gcusp_0&=4C_F\,, \\
\gcusp_1&=C_A C_F \left( \frac{268}{9} -8\zeta_2 \right) -C_F n_f \frac{40}{9}\,.
\end{align}

Since we have set up our factorization as a marginalization over $\vec k_\perp$, at intermediate stages of our calculation we will encounter vector plus distributions.
Definitions of vector plus distributions can be found in \cite{Chiu:2012ir,Ebert:2016gcn}. In particular, we will use the logarithmic plus distributions
\begin{align}
\cL_n(\vec p_\perp , \mu) \equiv \frac{1}{\pi \mu^2} \left[  \frac{\mu^2}{\vec p_\perp^{~2}} \ln^n \frac{\vec p_\perp^{~2}}{\mu^2}  \right]^\mu_+\,.
\end{align}
Since we will be interested in extracting the fixed order expansion of our resummed result, we will choose particular $\mu$ and $\nu$ scales. After having done this, all logarithms will appear in the form
\begin{align}
L^n_b\equiv\ln^n \left(  \frac{\vec b_\perp^{~2} Q^2 e^{2\gamma_E}}{4} \right)\,.
\end{align}
Relevant results for Fourier transforms of logarithms can be found in \cite{Ebert:2016gcn}. Here we will explicitly need 
\begin{align}
\text{FT}^{-1}[1]&= \delta(\vec p_\perp)\,,  \\
\text{FT}^{-1}[L_b]&= -\cL_0(\vec p_\perp,Q) \,, \\
\text{FT}^{-1}[L_b^2]&=2\cL_1 (\vec p_\perp,Q) \,, \\
\text{FT}^{-1}[L_b^3]&= -3\cL_2(\vec p_\perp,Q)-4\zeta_3 \delta(\vec p_\perp) \,, \\
\text{FT}^{-1}[L_b^4]&=4\cL_3(\vec p_\perp,Q)+16 \zeta_3 \cL_0(\vec p_\perp,Q) \,.
\end{align}

We can now show that our result reproduces the known leading power results for the EEC observable. We will expand the cross section perturbatively as 
\begin{align}
\frac{1}{\sigma_0}\frac{d\sigma}{dz}=\frac{d\sigma^{(0)}}{dz} +\frac{d\sigma^{(1)}}{dz} + \frac{d\sigma^{(2)}}{dz} +\cdots,
\end{align}
where the superscript indicates the perturbative order.
At LO and in the back-to-back region, we have
\begin{align}
\frac{d\sigma^{(0)}}{dz}&=\frac{1}{2} H^{(0)}(Q)  \int d^2 \vec k_\perp \int \frac{d^2 \vec b_\perp}{(2 \pi)^2} e^{-i \vec b_\perp \cdot \vec k_\perp}   \delta \left( 1-z- \frac{\vec k_\perp^2}{Q^2} \right ) \nn \\
&= \frac{1}{2} H^{(0)}(Q) \delta(1-z)\,.
\end{align}
Note that we have ignored the collinear region, which gives a
$\frac{1}{2}\delta(z)$ at LO.
To reproduce the NLO and NNLO fixed order results,  we choose to evaluate everything at the jet scale
\begin{align}
\mu_J^2=\frac{b_0^2}{\vec b_\perp^2}\,, \qquad \nu_J=Q\,.
\end{align}
This is convenient, since this is the natural $\mu$ scale for both the soft and jet functions.
We then have 
\begin{align}
L_H \to -L_b\,, \qquad L_r\to L_b\,, \qquad L_\perp \to 0\,.
\end{align}
This considerably simplifies the expression for the soft function
\begin{align}
S_\EEC\left(\vec b_\perp,\mu^2=\frac{b_0^2}{\vec b_\perp^2},\nu=Q\right) &=1+\left( \frac{\alpha_s(\mu_J)}{4\pi} \right) \left[ 
c^{\EEC}_1+\ugr_0 L_b
\right] +
\left(\frac{\alpha_s(\mu_J)}{4\pi}\right)^2 \left[
c^{\EEC}_2+\ugr_1 L_b
\right]\nn \\
&+\frac{1}{2}\left( \frac{\alpha_s(\mu_J)}{4\pi} \right)^2 \left[ 
c^{\EEC}_1+\ugr_0 L_b 
\right]^2 +\cO(\alpha_s^3)\,.
\end{align}
For the hard function, we have
\begin{align}
H\left( \mu^2=\frac{b_0^2}{\vec b_\perp^2}\right)&=1+ \left(\frac{\alpha_s(\mu_J)}{4\pi} \right)  \left(  -\frac{1}{2}\gcusp_0 L_b^2 -\gamma_0^H L_b +c_1^H  \right) \\
&+\left(\frac{\alpha_s (\mu_J)}{4\pi} \right)^2 \left[  \frac{1}{8} (\gcusp_0)^2 L_b^4 +\left( \frac{\beta_0 \gcusp_0}{6} +\frac{\gamma_0^H \gcusp_0}{2}  \right) L_b^3  \right. \nn \\
&\left. +\left(  \frac{(\gamma_0^H)^2}{2} +\frac{\beta_0 \gamma_0^H}{2} -\frac{\gcusp_1}{2}\right) L_b^2 -\gamma_1^H L_b +c_1^H \left( -\frac{\gcusp_0}{2}L_b^2 -L_b(\beta_0+\gamma_0^H)   \right) +c_2^H \right ]\nn\\
&+\cO(\alpha_s^3)\,. \nn
\end{align}
Finally, the one-loop constant for the jet function is
\begin{align}
J^{q}_\EEC\left (\vec b_\perp,\mu^2=\frac{b_0^2}{\vec b_\perp^2},\nu=Q \right ) = 1+\left( \frac{\alpha_s(\mu_J)}{4\pi}\right)c_1^J+ \cO(\alpha_s^2)\,,
\end{align}
where the term at $\cO(\alpha_s^2)$ is purely a constant with no logarithmic dependence, due to the choice of scales.

At NLO, we find
\begin{align}
\frac{d\sigma^{(1)}}{dz}&= \frac{1}{2} H^{(0)}(Q) \left( \frac{\alpha_s}{4\pi}  \right) \int d^2 \vec k_\perp \int \frac{d^2 \vec b_\perp}{(2 \pi)^2} e^{-i \vec b_\perp \cdot \vec k_\perp}  \left[ (c_1^H+2c_1^J+c_1^\EEC) -\gamma_0^HL_b -\frac{1}{2} \gcusp_0 L_b^2   \right]  \delta \left( 1-z- \frac{\vec k_\perp^2}{Q^2} \right ) \nn \\
&= \frac{1}{2} H^{(0)}(Q) \left( \frac{\alpha_s}{4\pi}  \right)  \int d^2 \vec k_\perp \int \frac{d^2 \vec b_\perp}{(2 \pi)^2}e^{-i \vec b_\perp \cdot \vec k_\perp} C_F \left[ (-4\zeta_2-8) +6L_b -2L_b^2   \right]  \delta \left( 1-z- \frac{\vec k_\perp^2}{Q^2} \right )  \nn \\
&=  \frac{1}{2} H^{(0)}(Q)\left( \frac{\alpha_s}{4\pi}  \right) C_F  \int d^2 \vec k_\perp \left[  (-4\zeta_2-8) \delta^{(2)}(\vec k_\perp)- 6 \cL_0(\vec k_\perp,\mu) -4 \cL_1(\vec k_\perp,\mu)  \right] \delta \left( 1-z- \frac{\vec k_\perp^2}{Q^2} \right ) \nn \\
&=  H^{(0)}(Q)\left( \frac{\alpha_s}{4\pi}  \right) \left(   C_F(-2\zeta_2-4)  \delta(1-z)-3C_F \left[ \frac{1}{1-z} \right]_+ -2C_F \left[ \frac{\ln(1-z)}{1-z} \right]_+ \right)\,.
\end{align}
To perform the final integral over the $\vec k_\perp$ appearing in the factorization theorem, we used
\begin{align}
&\int d^2 \vec k_\perp \cL_n (\vec k_\perp, Q) \delta \left(  1-z-\frac{\vec k_\perp^2}{Q^2}  \right) \nn \\
&\hspace{4cm}=\pi Q^2 \int d |\vec k_\perp | \delta \left(|\vec k_\perp| -Q \sqrt{1-z} \right) \cL_n (\vec k_\perp, Q) =\cL_n (1-z)\,,
\end{align}
where 
\begin{align}
\cL_n(1-z)=\left[ \frac{\ln(1-z)}{1-z}   \right]_+\,,
\end{align}
is the standard one-dimensional logarithmic plus distribution.

For the NNLO result, we provide slightly more details of the calculation. We begin by expanding the result in impact parameter space, keeping only the logarithmic terms. We find
\begin{align}
&\left.H\left( \mu^2=\frac{b_0^2}{\vec b_\perp^2}\right) \left[ J^{q}_\EEC \left(\vec b_\perp,\mu^2=\frac{b_0^2}{\vec b_\perp^2},\nu=Q\right) \right]^2 S_\EEC\left(\vec b_\perp,\mu^2=\frac{b_0^2}{\vec b_\perp^2},\nu=Q\right)\right|_{\alpha_s^2}\nn\\
&=\left(\frac{\alpha_s(\mu_J)}{4\pi}\right)^2 \frac{1}{8} (\gcusp_0)^2 L_b^4\nn \\
&+\left(\frac{\alpha_s(\mu_J)}{4\pi}\right)^2      \left[  \left(\frac{\beta_0 \gcusp_0}{6}+\frac{\gamma_0^H \gcusp_0}{2}   \right)-\frac{1}{2}\gcusp_0 \ugr_0  \right]        L_b^3 \nn \\
&+\left(\frac{\alpha_s(\mu_J)}{4\pi}\right)^2    \left[  \left( \frac{(\ugr_0)^2}{2}+ \frac{(\gamma_0^H)^2}{2}+\frac{\beta_0 \gamma_0^H}{2}-\left(\frac{\gcusp_1}{2}\right)  \right) +(c_1^\EEC+2c_1^J+ c_1^H) \left(-\frac{\gcusp_0}{2}\right) -\gamma_0^H \ugr_0                 \right]        L_b^2 \nn \\
&+\left(\frac{\alpha_s(\mu_J)}{4\pi}\right)^2    \left[   \ugr_1-\gamma_1^H-c_1^H(\beta_0+\gamma_0^H)+2c_1^J \ugr_0-2c_1^J\gamma_0^H-\gamma_0^H c_1^\EEC +\ugr_0 c_1^H          \right]        L_b\,.
\end{align}
In the literature, it is conventional to write the above expression evaluated with $\alpha_s$ at the hard scale, $Q$, which can be done using
\begin{align}
\alpha_s(\mu_J)=\alpha_s(Q) \left(  1+\frac{\alpha_s(Q)}{4\pi} \beta_0 L_b  \right)\,.
\end{align}
This modifies the NNLO result by $\beta_0$ terms multiplying the NLO result, namely by
\begin{align}
\left(\frac{\alpha_s(Q)}{4\pi}\right)^2 \left[ -\frac{1}{2} \gcusp_0 \beta_0 L_b^3 -\gamma_0^H \beta_0 L_b^2 +(2c_1^J +c_1^H +c_1^\EEC)\beta_0 L_b   \right]\,.
\end{align}
Written with $\alpha_s$ at the hard scale, we then have
\begin{align}
&H\cdot (J^q_\EEC(\vec b_\perp))^2 \cdot S_\EEC(\vec b_\perp)=\left(\frac{\alpha_s(Q)}{4\pi}\right)^2 \frac{1}{8} (\gcusp_0)^2 L_b^4 \\
&+\left(\frac{\alpha_s(Q)}{4\pi}\right)^2      \left[  -\frac{\beta_0 }{3}+\frac{\gamma_0^H }{2}  -\frac{\ugr_0 }{2}  \right]  \gcusp_0      L_b^3 \nn \\
&+\left(\frac{\alpha_s(Q)}{4\pi}\right)^2    \left[  \frac{(\ugr_0)^2}{2} + \frac{(\gamma_0^H)^2}{2}-\frac{\beta_0 \gamma_0^H}{2}-\frac{\gcusp_1}{2}-\gamma_0^H \ugr_0  -\frac{\gcusp_0}{2} (c_1^\EEC+2c_1^J+ c_1^H)                  \right]        L_b^2 \nn \\
&+\left(\frac{\alpha_s(Q)}{4\pi}\right)^2    \left[   \ugr_1-\gamma_1^H-c_1^H\gamma_0^H+2c_1^J \ugr_0-2c_1^J\gamma_0^H-\gamma_0^H c_1^\EEC +\ugr_0 c_1^H +(2c_1^J +c_1^\EEC)\beta_0          \right]        L_b\,. \nn
\end{align}
Performing the Fourier transform, we find
\begin{align} 
&\int d^2 \vec b_\perp e^{-i \vec b_\perp \cdot \vec k_\perp}H\cdot (J^q_\EEC(\vec b_\perp))^2 \cdot S_\EEC(\vec b_\perp)=\nn\\
&\left(\frac{\alpha_s(Q)}{4\pi}\right)^2 \frac{1}{8} (\gcusp_0)^2 (4\cL_3(\vec k_\perp, Q)+16\zeta_3 \cL_0(\vec k_\perp, Q)) \\
&+\left(\frac{\alpha_s(Q)}{4\pi}\right)^2       \left[  -\frac{\beta_0 }{3}+\frac{\gamma_0^H }{2}  -\frac{\ugr_0 }{2}  \right]  \gcusp_0           (-3\cL_2(\vec k_\perp, Q)) \nn \\
&+\left(\frac{\alpha_s(Q)}{4\pi}\right)^2     \left[  \frac{(\ugr_0)^2}{2} + \frac{(\gamma_0^H)^2}{2}-\frac{\beta_0 \gamma_0^H}{2}-\frac{\gcusp_1}{2}-\gamma_0^H \ugr_0  -\frac{\gcusp_0}{2} (c_1^\EEC+2c_1^J+ c_1^H)                  \right]       (2\cL_1(\vec k_\perp, Q)) \nn \\
&+\left(\frac{\alpha_s(Q)}{4\pi}\right)^2     \left[   \ugr_1-\gamma_1^H-c_1^H\gamma_0^H+2c_1^J \ugr_0-2c_1^J\gamma_0^H-\gamma_0^H c_1^\EEC +\ugr_0 c_1^H +(2c_1^J +c_1^\EEC)\beta_0          \right]        (-\cL_0(\vec k_\perp, Q))\,. \nn
\end{align}
This allows us to immediately write down the final result for the cross section in terms of the $z$ variable
\begin{align} \label{eq:final_NNLO}
&\frac{d\sigma^{(2)}}{dz}=\frac{1}{2} \left(\frac{\alpha_s(Q)}{4\pi}\right)^2 \frac{1}{2} (\gcusp_0)^2 \cL_3(1-z) \\
&+\frac{1}{2} \left(\frac{\alpha_s(Q)}{4\pi}\right)^2       \left[  \beta_0-\frac{3\gamma_0^H }{2}  +\frac{3\ugr_0 }{2}  \right]  \gcusp_0           \cL_2(1-z) \nn \\
&+\frac{1}{2} \left(\frac{\alpha_s(Q)}{4\pi}\right)^2     \left[  (\ugr_0)^2 + (\gamma_0^H)^2-\beta_0 \gamma_0^H-\gcusp_1-2\gamma_0^H \ugr_0  -\gcusp_0 (c_1^\EEC+2c_1^J+ c_1^H)                  \right]       \cL_1(1-z) \nn \\
&+\frac{1}{2} \left(\frac{\alpha_s(Q)}{4\pi}\right)^2     \left[  2 (\gcusp_0)^2\zeta_3- \ugr_1+\gamma_1^H \right. \nn \\
& \left. \hspace{3cm}+c_1^H\gamma_0^H-2c_1^J \ugr_0+2c_1^J\gamma_0^H+\gamma_0^H c_1^\EEC -\ugr_0 c_1^H -(2c_1^J +c_1^\EEC)\beta_0          \right]        \cL_0(1-z)\,. \nn
\end{align}
Plugging in the values of the different anomalous dimensions, we have
\begin{align} 
\frac{d\sigma^{(2)}}{dz}&=\left(\frac{\alpha_s(Q)}{4\pi}\right)^2 4C_F^2 \cL_3(1-z) \\
&+\left(\frac{\alpha_s(Q)}{4\pi}\right)^2       \left[  18 C_F^2 +\frac{22}{3}C_A C_F -\frac{4}{3}n_f C_F\right]           \cL_2(1-z) \nn \\
&+\left(\frac{\alpha_s(Q)}{4\pi}\right)^2     \left[ C_F C_A(4 \zeta_2
  - \frac{35}{9}) + \frac{2}{9} C_F n_f + C_F^2 (8 \zeta_2+  34)  \right]       \cL_1(1-z) \nn \\
&+\left(\frac{\alpha_s(Q)}{4\pi}\right)^2     \left[  (\frac{45}{2}+24\zeta_2-8\zeta_3) C_F^2 + (-\frac{35}{2}+22\zeta_2+12\zeta_3)  C_FC_A + (3-4\zeta_2) C_F n_f     \right]        \cL_0(1-z)\,. \nn
\end{align}
We can compare this result to a previous NNLL result in the literature \cite{deFlorian:2004mp}, computed in the CSS formalism.
The result of \cite{deFlorian:2004mp} was written as
\begin{align}\label{eq:CSS}
\frac{1}{\sigma_T} \frac{d\sigma}{d\cos \chi} &=\frac{1}{4y} \frac{\alpha_s(Q)}{\pi} \left[-A^{(1)} \ln y +B^{(1)}     \right] \nn \\
&+\frac{1}{4y} \left( \frac{\alpha_s(Q)}{\pi}\right)^2 \left[  \frac{1}{2} (A^{(1)})^2 \ln^3 y +\left(-\frac{3}{2} B^{(1)} A^{(1)} +\frac{\beta_0}{4} A^{(1)}    \right) \ln^2 y  \right. \nn \\
&\hspace{3cm} +\left(  -A^{(2)} -\frac{\beta_0}{4} B^{(1)} + (B^{(1)})^2 -A^{(1)}H^{(1)} \right) \ln y    \nn \\
&\hspace{3cm}\left.  +B^{(2)} +B^{(1)}H^{(1)} +2\zeta_3  (A^{(1)})^2    \vphantom{\frac{1}{2}}     \right]\,,
\end{align}
where $y=\sin^2 (\pi - \chi)/2 = 1 - z$.
The required constants appearing in \Eq{eq:CSS} are given by
\begin{align}
A^{(1)}&=\frac{\gcusp_0}{4}\,, \qquad
A^{(2)}=\frac{\gcusp_1}{16}\,, \\
B^{(1)}&=-\frac{3}{2}C_F\,, \qquad
B^{(2)}=-\frac{1}{2}\gamma_q^{(2)} +C_F \frac{\beta_0}{4} \left( 5\zeta_2  -2 \right)\,, \\
H^{(1)}&=-C_F \left(   \frac{11}{4}+\zeta_2 \right)\,,
\end{align}
and
\begin{align}
\gamma_q^{(2)}=C_F^2 \left(   \frac{3}{8} -3\zeta_2+6\zeta_3\right) +C_F C_A \left(  \frac{17}{24} +\frac{11}{3}\zeta_2 -3\zeta_3  \right) -C_F n_f T_R \left( \frac{1}{6} +\frac{4}{3}\zeta_2   \right)\,.
\end{align}
Note that to perform the comparison, one must take into account that the formula of \Eq{eq:CSS} from \cite{deFlorian:2004mp} normalizes to the NLO total cross section
\begin{align}
\sigma_T=\sigma_0 \left( 1+3\left(\frac{\alpha_s(Q)}{4\pi}\right) C_F \right)\,,
\end{align}
while in \Eq{eq:final_NNLO}, we have normalized only to $\sigma_0$. We find exact agreement with their result. The result of \Eq{eq:CSS} was verified by comparison with the fixed order program \textsc{Event2} \cite{Catani:1996jh,Catani:1996vz}, and was shown to correctly reproduce the logarithmic structure to this order. This provides a highly non-trivial check of our factorization theorem. In particular, the difference found in \cite{deFlorian:2004mp} between the $B^{(2)}$ coefficient for the EEC and $B^{(2)}_{q,\text{NS}}$ (see equations 19 and 21 in \cite{deFlorian:2004mp}) is naturally reproduced by our factorization theorem.

\bibliography{EEC_bib}{}
\bibliographystyle{JHEP}

\end{document}